\documentclass[12pt]{article}
\usepackage{a4wide}
\usepackage{latexsym}
\usepackage{amsmath}
\usepackage{amsfonts}
\usepackage{amssymb}
\usepackage{amscd}
\usepackage{cite}
\usepackage{placeins}

\usepackage{pslatex}
\usepackage{graphicx}
\usepackage[latin1,utf8]{inputenc}
\usepackage[T1]{fontenc}

\allowdisplaybreaks

\newcommand{\bq}{\begin{eqnarray}}
\newcommand{\eq}{\end{eqnarray}}

\newcommand{\slashoperator}[2]{|_{#2} #1}

\begin{document}

\thispagestyle{empty}

\begin{flushright}
  MITP/20-057
\end{flushright}

\vspace{1.5cm}

\begin{center}
  {\Large\bf Numerical evaluation of iterated integrals related to elliptic Feynman integrals\\
  }
  \vspace{1cm}
  {\large Moritz Walden and Stefan Weinzierl \\
  \vspace{1cm}
      {\small \em PRISMA Cluster of Excellence, Institut f{\"u}r Physik, }\\
      {\small \em Johannes Gutenberg-Universit{\"a}t Mainz,}\\
      {\small \em D - 55099 Mainz, Germany}\\
  } 
\end{center}

\vspace{2cm}

\begin{abstract}\noindent
  {
We report on an implementation within GiNaC to evaluate iterated integrals related to elliptic Feynman
integrals numerically to arbitrary precision within the region of convergence of the series expansion of the integrand.
The implementation includes iterated integrals of modular forms as well as
iterated integrals involving the Kronecker coefficient functions $g^{(k)}(z,\tau)$.
For the Kronecker coefficient functions iterated integrals in $d\tau$ and $dz$ are implemented.
This includes elliptic multiple polylogarithms.
   }
\end{abstract}

\vspace*{\fill}

\newpage

\section{Introduction}
\label{sect:intro}

Certain Feynman integrals evaluate to transcendental functions related to elliptic curves, 
or slightly more general to transcendental functions related to the moduli space ${\mathcal M}_{1,n}$ of
a genus one curve with $n$ marked points.
These functions are beyond the class of multiple polylogarithms, which can be viewed as
transcendental functions related to the moduli space ${\mathcal M}_{0,n}$ of
a genus zero curve with $n$ marked points.
Let us call these Feynman integrals ``elliptic Feynman integrals''.
The elliptic Feynman integrals have received considerable attention 
in recent years \cite{Broadhurst:1993mw,Laporta:2004rb,Bailey:2008ib,MullerStach:2011ru,Adams:2013nia,Bloch:2013tra,Remiddi:2013joa,Adams:2014vja,Adams:2015gva,Adams:2015ydq,Bloch:2016izu,Adams:2017ejb,Bogner:2017vim,Adams:2018yfj,Honemann:2018mrb,Bloch:2014qca,Sogaard:2014jla,Tancredi:2015pta,Primo:2016ebd,Remiddi:2016gno,Adams:2016xah,Bonciani:2016qxi,vonManteuffel:2017hms,Adams:2017tga,Ablinger:2017bjx,Primo:2017ipr,Passarino:2017EPJC,Remiddi:2017har,Bourjaily:2017bsb,Hidding:2017jkk,Broedel:2017kkb,Broedel:2017siw,Broedel:2018iwv,Lee:2017qql,Lee:2018ojn,Adams:2018bsn,Adams:2018kez,Broedel:2018qkq,Bourjaily:2018yfy,Bourjaily:2018aeq,Besier:2018jen,Mastrolia:2018uzb,Ablinger:2018zwz,Frellesvig:2019kgj,Broedel:2019hyg,Blumlein:2019svg,Broedel:2019tlz,Bogner:2019lfa,Kniehl:2019vwr,Broedel:2019kmn,Abreu:2019fgk,Duhr:2019rrs,2019arXiv190811815L,Klemm:2019dbm,Bonisch:2020qmm}.
It is worth mentioning that similar transcendental functions also occur in string theory \cite{Broedel:2014vla,Broedel:2015hia,Broedel:2017jdo,DHoker:2015wxz,Hohenegger:2017kqy,Broedel:2018izr}.

It is therefore desirable to have numerical evaluation routines for these transcendental functions.
For multiple polylogarithms, the numerical evaluation routines within the GiNaC library \cite{Bauer:2000cp} 
are widely used \cite{Vollinga:2004sn}.
In this paper we report on an implementation
within GiNaC to evaluate transcendental functions related to elliptic Feynman
integrals numerically to arbitrary precision within the region of convergence.
The GiNaC library is open source software and freely available.

The possibility to evaluate these transcendental functions to arbitrary precision is useful for the PSLQ algorithm \cite{Ferguson:1992}.
In the context of Feynman integrals, the PSLQ algorithm is often 
employed to fix boundary constants, if the Feynman integrals are calculated from their differential equations.

In the literature there exist various slightly different definitions of transcendental functions related to elliptic Feynman integrals.
For the numerical evaluation we support a wide range of these.
Common to all definitions is the fact that the transcendental functions are defined as iterated integrals on a covering space
of the moduli space ${\mathcal M}_{1,n}$.
Standard coordinates on this space are $(z_1,\dots,z_{n-1},\tau)$, where $z_1, \dots, z_{n-1}$ denote the positions of $(n-1)$ punctures and
$\tau$ describes the shape of the torus.
Due to translational invariance we may assume that one puncture is at the origin: $z_0=0$.
From these standard coordinates we may already divide the transcendental functions into two broad classes:
The first class consists of iterated integrals, where the integration variable is $\tau$,
the second class consists of iterated integrals, where the integration variable is a $z$ variable.
The former class includes for example iterated integrals of modular forms, the latter class the $\tilde{\Gamma}$ functions,
also known as (meromorphic, but non-double periodic) elliptic multiple polylogarithms.

Supporting a wide variety of integrands, which can be combined in any reasonable fashion, has one limitation:
We only support the numerical evaluation in regions, where all integrands have a convergent Laurent series expansion with at most
a simple pole at the base point of the integration.

This paper is organised as follows:
In section~\ref{sect:definitions} we introduce our notation, the necessary background on iterated integrals
and the relevant special functions, which are later used to define the various integrands.
In section~\ref{sect:kernels} we discuss the various integrands or integration kernels for the iterated integrals.
Section~\ref{sect:implementation} describes the implementation within the GiNaC library.
In section~\ref{sect:examples} we give several examples on how to use our routines.
In section~\ref{sect:advanced} we discuss advanced usage, limitations and give an outlook, how the performance for specific sub-classes
of iterated integrals can be improved.
Finally, section~\ref{sect:conclusions} contains our conclusions.

Appendix~\ref{appendix:standard_functions} summarises our notation for some standard mathematical functions.
Appendix~\ref{appendix:agm} is devoted to the arithmetic-geometric mean, which is used to compute numerically complete elliptic integrals of the first and
second kind.
Appendix~\ref{appendix:Kronecker_symbol} gives all relevant details on the Kronecker symbol.


\section{Definitions}
\label{sect:definitions}

\subsection{Notation}
\label{sect:notation}

In the literature one finds that both the variable $\exp(\pi i \tau )$
as well as the variable $\exp(2 \pi i \tau )$ are used.
To avoid confusion we will use throughout this paper the notation
\bq
\label{def_basic_variables}
 q \; =\; \exp\left(\pi i \tau \right),
 & &
 \bar{q} \; = \; \exp\left(2 \pi i \tau \right),
 \nonumber \\
 w \; =\; \exp\left(\pi i z \right),
 & &
 \bar{w} \; = \; \exp\left(2 \pi i z \right).
\eq
Most of our formula will be in barred variables.
The barred variables are periodic with period $1$, e.g. $\bar{q}(\tau+1)=\bar{q}(\tau)$.


\subsection{Iterated integrals}
\label{sect:iterated_integrals}

Let us start with the general definition of an iterated integrals \cite{Chen}:
Let $M$ be a $n$-dimensional (complex) manifold and
\bq
 \gamma & : & \left[a,b\right] \rightarrow M
\eq
a path with start point ${x}_i=\gamma(a)$ and end point ${x}_f=\gamma(b)$.
Suppose further that $\omega_1$, ..., $\omega_r$ are differential $1$-forms on $M$.
Let us write
\bq
 f_j\left(\lambda\right) d\lambda & = & \gamma^\ast \omega_j
\eq
for the pull-backs to the interval $[a,b]$.
For $\lambda \in [a,b]$ the $k$-fold iterated integral 
of $\omega_1$, ..., $\omega_r$ along the path $\gamma$ is defined
by
\bq
 I_{\gamma}\left(\omega_1,...,\omega_r;\lambda\right)
 & = &
 \int\limits_a^{\lambda} d\lambda_1 f_1\left(\lambda_1\right)
 \int\limits_a^{\lambda_1} d\lambda_2 f_2\left(\lambda_2\right)
 ...
 \int\limits_a^{\lambda_{r-1}} d\lambda_r f_r\left(\lambda_r\right).
\eq
We define the $0$-fold iterated integral to be
\bq
 I_{\gamma}\left(;\lambda\right)
 & = &
 1.
\eq
Our main applications will be the cases where $\dim_{\mathbb C} M=1$ with coordinate $z$.
Our standard integration path in the absence of trailing zeros
will be the line segment from zero to $z_0 \in {\mathbb C}$.
The case of trailing zeros will be discussed in the next paragraph.
We take $|z|$ as curve parameter and hence $[a,b]=[0,\lambda_0]$ with $\lambda_0=|z_0|$.
In the sequel we drop in the notation of the iterated integral the dependence on the integration path $\gamma$
and simply write $I(\omega_1,...,\omega_r;z_0)$.

\subsubsection{Shuffle product and trailing zeros}

Let $\lambda_0 \in {\mathbb R}_{>0}$ and denote by $U$ the domain $U=\{z\in{\mathbb C} | |z| \le \lambda_0 \}$.
Let us assume that all $\omega_j$ are holomorphic in $U\backslash\{0\}$ and have at most a simple pole
at $z=0$.
In other words
\bq
 \omega_j
 & = &
 f_j\left(z\right) dz
 \; = \;
 \sum\limits_{n=0}^\infty c_{j,n} \; z^{n-1} dz,
 \;\;\;\;\;\;\;\;\;
 c_{j,n} \; \in \; {\mathbb C}.
\eq
We say that $\omega_j$ has a trailing zero, if $c_{j,0} \neq 0$.
We denote by
\bq
\label{def_L0}
 L_0 & = & l_0(z) dz \;\; = \;\; \frac{dz}{z}
\eq
the logarithmic form with $c_{0}=1$ and $c_n=0$ for $n>0$.
We extend the definition of iterated integrals: We set
\bq
\label{only_trailing_zeros}
 I(\underbrace{L_0,\dots,L_0}_{r};z_0)
 & = &
 \frac{1}{r!} \ln^r\left(z_0\right)
\eq
and define recursively
\bq
 I\left(\omega_1,\omega_2,\dots,\omega_r;z_0\right)
 & = &
 \int\limits_0^{z_0} dz_1 f_1\left(z_1\right)
 I\left(\omega_2,\dots,\omega_r;z_1\right).
\eq
This agrees with our previous definition in the case that $\omega_r$ has no trailing zero, but allows
for trailing zeros.
We say that the iterated integral $I(\omega_1,\dots,\omega_r;z_0)$ has a trailing zero,
if $\omega_r$ has a trailing zero.
If $\omega_r$ has a trailing zero, we may always write
\bq
\label{def_regularised_form}
 \omega_r
 & = &
 c_{r,0} L_0 + \omega_r^{\mathrm{reg}},
\eq
with
\bq
 \omega_r^{\mathrm{reg}}
 & = &
 \sum\limits_{n=1}^\infty c_{j,n} \; z^{n-1} dz
\eq
having no trailing zero.

Iterated integrals come with a shuffle product:
\bq
 I(\omega_{1},\dots,\omega_{k};z_0)
 \cdot
 I(\omega_{k+1},\dots,\omega_{r};z_0)
 & = &
 \sum\limits_{\mathrm{shuffles} \; \sigma}
 I(\omega_{\sigma(1)},\dots,\omega_{\sigma(r)};z_0),
\eq
where the sum runs over all shuffles $\sigma$ of $(1,\dots,k)$ with $(k+1,\dots,r)$.
A shuffle is a permutation of $(1,\dots,r)$, which preserves the relative order of 
$(1,\dots,k)$ and $(k+1,\dots,r)$.
We may use the shuffle product and eq.~(\ref{def_regularised_form}) to remove trailing zeros,
for example if $c_{1,0}=0$ and $c_{2,0}=1$ we have
\bq
 I(\omega_{1},\omega_{2};z_0)
 & = &
 I(\omega_{1},L_0;z_0)
 +
 I(\omega_{1},\omega_{2}^{\mathrm{reg}};z_0)
 \nonumber \\
 & = &
 I(L_0;z_0) I(\omega_{1};z_0)
 -
 I(L_0,\omega_{1};z_0)
 +
 I(\omega_{1},\omega_{2}^{\mathrm{reg}};z_0).
\eq
This isolates all trailing zeros in integrals of the form~(\ref{only_trailing_zeros}),
for which we may use the explicit formula in eq.~(\ref{only_trailing_zeros}).
It is therefore sufficient to focus on iterated integrals with no trailing zeros.
For
\bq
 I(\omega_{1},\dots,\omega_{r};z_0)
\eq
this means $c_{r,0}=0$.
Please note that $c_{k,0} \neq 0$ is allowed for $k<r$
and in particular that the form $L_0$ is allowed in positions $k<r$.

For integrals with no trailing zeros we introduce the notation
\bq
 I_{m_1,\dots,m_r}(\omega_{1},\dots,\omega_{r};z_0)
 & = &
 I(\underbrace{L_0,\dots,L_0}_{m_1-1},\omega_{1},\dots,\omega_{r-1},\underbrace{L_0,\dots,L_0}_{m_r-1},\omega_{r};z_0),
\eq
where we assumed that $\omega_k \neq L_0$ and $(m_k-1)$ $L_0$'s precede $\omega_k$.
This notation resembles the notation of multiple polylogarithms.
The motivation for this notation is as follows:
The iterated integrals $I_{m_1,\dots,m_r}(\omega_{1},\dots,\omega_{r};z_0)$ have just a $r$-fold series expansion,
and not a $(m_1+\dots+m_r)$-fold one.

\subsubsection{Series expansion}

With the same assumptions as in the previous subsection
(all $\omega_j$ are holomorphic in $U\backslash\{0\}$ and have at most a simple pole at $z=0$)
an iterated integral with no trailing zero has a convergent series expansion
in $U$:
\bq
\label{iter_int_series_expansion}
 I_{m_1,\dots,m_r}(\omega_{1},\dots,\omega_{r};z_0)
 & = &
 \sum\limits_{i_1=1}^\infty \sum\limits_{i_2=1}^{i_1} \dots \sum\limits_{i_r=1}^{i_{r-1}}
  z_0^{i_1}
  \frac{c_{1,i_1-i_2} \dots c_{r-1,i_{r-1}-i_r} c_{r,i_r}}
       {i_1^{m_1} i_2^{m_2} \cdot \dots \cdot i_r^{m_r}}.
\eq
This formula can be used for the numerical evaluation of the iterated integral:
We truncate the outer sum over at $i_1=N$. 
Let us write eq.~(\ref{iter_int_series_expansion}) as
\bq
 I_{m_1,\dots,m_r}(\omega_{1},\dots,\omega_{r};z_0)
 & = &
 \sum\limits_{i_1=1}^\infty d_{i_1},
 \nonumber \\
 d_{i_1} & = &
  z_0^{i_1}
  \sum\limits_{i_2=1}^{i_1} \dots \sum\limits_{i_r=1}^{i_{r-1}}
  \frac{c_{1,i_1-i_2} \dots c_{r-1,i_{r-1}-i_r} c_{r,i_r}}
       {i_1^{m_1} i_2^{m_2} \cdot \dots \cdot i_r^{m_r}}.
\eq
This gives a numerical approximation $I^{\mathrm{approx}}(N)$ of the iterated integral
\bq
 I^{\mathrm{approx}}(N)
 & = &
 \sum\limits_{i_1=1}^N d_{i_1}.
\eq
Choosing $N$ large enough, such that the neglected terms contribute below the numerical precision
gives the numerical evaluation of the iterated integral.

In more detail, let us define for two numbers $a$ and $b$ an equivalence relation.
We say $a \sim b$, if they have exactly the same floating-point representation
within a given numerical precision.
Our standard truncation criterion is as follows: We truncate the iterated integral at $N$ if
\bq
\label{truncation_criterion}
 I^{\mathrm{approx}}\left(N\right)
 \; \sim \;
 I^{\mathrm{approx}}\left(N-1\right)
 & \mbox{and} &
 d_N \; \neq \; 0.
\eq
This gives reliable results in most cases.
However, there are a few specific cases where this criterion is inappropriate.
These specific cases are discussed in more detail in section~\ref{sect:advanced}.
In order to handle also these cases, we provide as an alternative method the truncation at a user-specified value
$N_{\mathrm{user}}$.

\subsection{Special functions}
\label{sect:special_functions}

In order to discuss elliptic multiple polylogarithms and related iterated integrals 
we first have to introduce a few special functions related to modular forms.

\subsubsection{Basics of modular forms}

We denote by ${\mathbb H}$ the complex upper half-plane with coordinate $\tau$:
\bq
 \mathbb{H}
 & = &
 \left\{ \tau \in \mathbb{C} | \mathrm{Im}(\tau) > 0 \right\}.
\eq
A modular transformation is given by
\bq
 \tau'
 \; = \;
 \frac{a \tau +b}{c \tau +d},
 & &
 \tau
 \; \in \;
 {\mathbb H}, 
 \;\;\;\;\;\;
 \left( \begin{array}{cc}
 a & b \\
 c & d \\
 \end{array} \right)
 \; \in \; \mathrm{SL}_2\left({\mathbb Z}\right).
\eq
A meromorphic function $f: \mathbb{H} \rightarrow \mathbb{C}$ is a modular form
of modular weight $k$ for $\mathrm{SL}_2(\mathbb{Z})$ if
\begin{enumerate}
\item $f$ transforms under modular transformations as
\bq
\label{trafo_modular_form}
 f\left( \dfrac{a\tau+b}{c\tau+d} \right) = (c\tau+d)^k \cdot f(\tau) 
 \qquad \text{for} \;\; \gamma = \left( \begin{array}{cc}
a & b \\ 
c & d
\end{array} \right) \in \mathrm{SL}_2(\mathbb{Z}),
\eq
\item $f$ is holomorphic on $\mathbb{H}$,
\item $f$ is holomorphic at $i \infty$.
\end{enumerate}
It is convenient to introduce the $\slashoperator{\gamma}{k}$ operator,
defined by
\bq
(f \slashoperator{\gamma}{k})(\tau) & = & (c\tau+d)^{-k} \cdot f(\gamma(\tau)).
\eq
With the help of the $\slashoperator{\gamma}{k}$ operator we may rewrite eq.~(\ref{trafo_modular_form}) as
$(f \slashoperator{\gamma}{k}) = f$ for $\gamma \in \mathrm{SL}_2(\mathbb{Z})$.
A modular form for $\mathrm{SL}_2(\mathbb{Z})$ is called a cusp form of $\mathrm{SL}_2(\mathbb{Z})$, if it vanishes at the cusp $\tau=i \infty$.

Apart from $\mathrm{SL}_2({\mathbb Z})$ we may also look at congruence subgroups.
The standard congruence subgroups are defined by
\begin{align}
 \Gamma_0(N) 
 & = 
 \left\{ \left( \begin{array}{cc}
                 a & b \\ 
                 c & d
                \end{array}  \right)
         \in \mathrm{SL}_2(\mathbb{Z}): c \equiv 0\ \text{mod}\ N
 \right\},
 \nonumber \\
 \Gamma_1(N) 
 & = 
 \left\{ \left( \begin{array}{cc}
                 a & b \\ 
                 c & d
                 \end{array}  \right)
         \in \mathrm{SL}_2(\mathbb{Z}): a,d \equiv 1\ \text{mod}\ N, \; c \equiv 0\ \text{mod}\ N
 \right\},
 \nonumber \\
 \Gamma(N) 
 & = 
 \left\{ \left( \begin{array}{cc}
                a & b \\ 
                c & d
                \end{array}  \right) \in \mathrm{SL}_2(\mathbb{Z}): a,d \equiv 1\ \text{mod}\ N, \; b,c \equiv 0\ \text{mod}\ N
 \right\}.
 \nonumber
\end{align}
$\Gamma(N)$ is called the principle congruence subgroup of level $N$.
The principle congruence subgroup $\Gamma(N)$ is a normal subgroup of $\mathrm{SL}_2({\mathbb Z})$.
In general, a subgroup $\Gamma$ of $\mathrm{SL}_2({\mathbb Z})$ is called a congruence subgroup,
if there exists an $N$ such that
\bq
 \Gamma\left(N\right) & \subseteq & \Gamma.
\eq
The smallest such $N$ is called the level of the congruence subgroup.

We may now define modular forms for a congruence subgroup $\Gamma$,
by relaxing the transformation law in eq.~(\ref{trafo_modular_form}) to hold only for
modular transformations from the subgroup $\Gamma$, plus holomorphicity on ${\mathbb H}$ and at the cusps.
In detail:
A meromorphic function $f: \mathbb{H} \rightarrow \mathbb{C}$ 
is a modular form of modular weight $k$ for the congruence subgroup $\Gamma$
if
\begin{enumerate}
\item $f$ transforms as
\bq
 (f \slashoperator{\gamma}{k}) 
 & = & 
 f
 \qquad \text{for} \;\; 
 \gamma 
 \in \Gamma,
\eq
\item $f$ is holomorphic on $\mathbb{H}$,
\item $f \slashoperator{\gamma}{k}$ is holomorphic at $i \infty$ for all $\gamma \in \mathrm{SL}_2({\mathbb Z})$.
\end{enumerate}
A modular form $f$ for a congruence subgroup $\Gamma$ is called a cusp form of $\Gamma$, if $f \slashoperator{\gamma}{k}$
vanishes at $\tau=i \infty$ for all $\gamma \in \mathrm{SL}_2({\mathbb Z})$.

For a congruence subgroup $\Gamma$ of $\mathrm{SL}_2({\mathbb Z})$ we denote by
${\mathcal M}_k(\Gamma)$ the space of modular forms of weight $k$ for $\Gamma$.
Furthermore, $\mathcal{M}_k(\Gamma)$ is the direct sum of two finite dimensional $\mathbb{C}$-vector spaces:
the space of cusp forms $ \mathcal{S}_k(\Gamma)$ and the Eisenstein subspace $\mathcal{E}_k(\Gamma)$.

\subsubsection{The Kronecker function}

We start with the Kronecker function $F(x,y,\tau)$.
This function is defined in terms of the first Jacobi theta function by
\bq
 F\left(x,y,\tau\right)
 & = &
 \pi
 \theta_1'\left(0,q\right) \frac{\theta_1\left( \pi\left(x+y\right), q \right)}{\theta_1\left( \pi x, q \right)\theta_1\left( \pi y, q \right)},
\eq
where $q=\exp(\pi i \tau)$.
The definition of the Jacobi theta function is given in appendix~\ref{appendix:standard_functions}
and $\theta_1'$ denotes the derivative with respect to the first argument.
It is obvious from the definition that the Kronecker function is symmetric in $x$ and $y$.
We are interested in the Laurent expansion in one of these variables.
We define functions
$g^{(k)}(z,\tau)$ through
\bq
\label{def_g_n}
 F\left(z,\alpha,\tau\right)
 & = &
 \sum\limits_{k=0}^\infty g^{(k)}\left(z,\tau\right) \alpha^{k-1}.
\eq
The functions $g^{(k)}(z,\tau)$ will enter the definition of elliptic multiple polylogarithms.
Let us recall some of their properties \cite{Brown:2011,Broedel:2018qkq}.
When viewed as a function of $z$, the function $g^{(k)}(z,\tau)$ has only simple poles.
More concretely, the function $g^{(1)}(z,\tau)$ has a simple pole with unit residue at every point of the lattice.
For $k>1$ the function $g^{(k)}(z,\tau)$ has a simple pole only at those lattice points 
that do not lie on the real axis.
The (quasi-) periodicity properties are
\bq
 g^{(k)}\left(z+1,\tau\right) & = &  g^{(k)}\left(z,\tau\right),
 \nonumber \\
 g^{(k)}\left(z+\tau,\tau\right) & = &  
 \sum\limits_{j=0}^k \frac{\left(-2\pi i\right)^j}{j!} g^{(k-j)}\left(z,\tau\right).
\eq
We see that $g^{(k)}(z,\tau)$ is invariant under translations by $1$, but not by $\tau$.
The functions $g^{(k)}(z,\tau)$ have the symmetry
\bq
 g^{(k)}(-z,\tau)
 & = &
 \left(-1\right)^k g^{(k)}(z,\tau).
\eq
In previous publications we introduced the notation \cite{Adams:2014vja,Adams:2015gva,Adams:2015ydq,Adams:2016xah,Bogner:2019lfa}
\bq
 \mathrm{ELi}_{n;m}\left(\bar{u};\bar{v};\bar{q}\right) & = & 
 \sum\limits_{j=1}^\infty \sum\limits_{k=1}^\infty \; \frac{\bar{u}^j}{j^n} \frac{\bar{v}^k}{k^m} \bar{q}^{j k}
\eq
and the linear combinations
\bq
 \overline{\mathrm{E}}_{n;m}\left(\bar{u};\bar{v};\bar{q}\right) 
 & = &
  \mathrm{ELi}_{n;m}\left(\bar{u};\bar{v};\bar{q}\right)
  - \left(-1\right)^{n+m} \mathrm{ELi}_{n;m}\left(\bar{u}^{-1};\bar{v}^{-1};\bar{q}\right).
\eq
For $\tau \in {\mathbb H}$ the function $\mathrm{ELi}_{n;m}(\bar{u};\bar{v};\bar{q})$
converges for 
\bq
 \left|\bar{u}\right|,
 \left|\bar{v}\right|
 & < & \left|\bar{q}\right|^{-1},
\eq
and the function $\overline{\mathrm{E}}_{n;m}\left(\bar{u};\bar{v};\bar{q}\right)$ converges for
\bq
\label{convergence_Ebar}
 \left|\bar{q}\right|
 \;\; < \;\; 
 \left|\bar{u}\right|,
 \left|\bar{v}\right|
 \;\; < \;\; 
 \left|\bar{q}\right|^{-1},
\eq
For 
\bq
 \bar{u} \; = \; \exp\left(2\pi i x \right),
 \;\;\;
 \bar{v} \; = \; \exp\left(2\pi i y \right),
 \;\;\;
 \bar{q} \; = \; \exp\left(2\pi i \tau \right)
\eq
eq.~(\ref{convergence_Ebar}) translates to
\bq
 - \mathrm{Im}\left(\tau\right)
 \;\; < \;\; 
 \mathrm{Im}\left(x\right),
 \mathrm{Im}\left(y\right)
 \;\; < \;\; 
 \mathrm{Im}\left(\tau\right).
\eq
The functions $\overline{\mathrm{E}}_{n;m}$ are helpful for the $\bar{q}$-expansion of the functions $g^{(k)}(z,\tau)$.
Explicitly one has with $\bar{q}=\exp(2\pi i\tau)$ and $\bar{w}=\exp(2\pi i z)$
\bq
\label{g_n_explicit}
 g^{(0)}\left(z,\tau\right)
 & = & 1,
 \nonumber \\
 g^{(1)}\left(z,\tau\right)
 & = &
 - 2 \pi i \left[
                  \frac{1+\bar{w}}{2 \left(1-\bar{w}\right)}
                  + \overline{\mathrm{E}}_{0,0}\left(\bar{w};1;\bar{q}\right)
 \right],
 \nonumber \\
 g^{(k)}\left(z,\tau\right)
 & = &
 - \frac{\left(2\pi i\right)^k}{\left(k-1\right)!} 
 \left[
 - \frac{B_k}{k}
       + \overline{\mathrm{E}}_{0,1-k}\left(\bar{w};1;\bar{q}\right)
 \right],
 \;\;\;\;\;\;\;\;\;\;\;\;\;\;\;\;\;\;\;\;\;\;\;\;\;\;\;
 k > 1,
\eq
where $B_k$ denote the $k$-th Bernoulli number, defined by
\bq
 \frac{x}{e^x-1}
 & = &
 \sum\limits_{k=0}^\infty \frac{B_k}{k!} x^k.
\eq
It will be convenient to set $g^{(-1)}(z,\tau)=0$.

Under modular transformations the functions $g^{(k)}(z,\tau)$ transform as
\bq
 g^{(k)}\left(\frac{z}{c\tau+d},\frac{a\tau+b}{c\tau+d}\right)
 & = &
 \left(c\tau +d \right)^k
 \sum\limits_{j=0}^k
 \frac{\left(2\pi i\right)^j}{j!}
 \left( \frac{c z}{c\tau+d} \right)^j
 g^{(k-j)}\left(z,\tau\right).
\eq

\subsubsection{Eisenstein series for $\mathrm{SL}_2({\mathbb Z})$}

The $z$-dependent Eisenstein series $E_k(z,\tau)$ are defined by
\bq
 E_k\left(z,\tau\right)
 & = &
 \sideset{}{_e}\sum\limits_{(n_1,n_2) \in {\mathbb Z}^2} \frac{1}{\left(z+n_1 + n_2\tau \right)^k}.
\eq
The series is absolutely convergent for $k \ge 3$.
For $k=1$ and $k=2$ the Eisenstein summation depends on the choice of generators. The Eisenstein summation prescription is defined by
\bq
 \sideset{}{_e}\sum\limits_{(n_1,n_2) \in {\mathbb Z}^2} f\left(z+n_1 + n_2\tau \right)
 & = &
 \lim\limits_{N_2\rightarrow \infty} \sum\limits_{n_2=-N_2}^{N_2}
 \left(
 \lim\limits_{N_1\rightarrow \infty} \sum\limits_{n_1=-N_1}^{N_1}
 f\left(z + n_1 + n_2 \tau \right)
 \right).
\eq
One further sets
\bq
\label{def_e_k}
 e_k\left(\tau\right)
 & = &
 \sideset{}{_e}\sum\limits_{(n_1,n_2) \in {\mathbb Z}^2\backslash (0,0)} \frac{1}{\left(n_1 + n_2\tau \right)^k}.
\eq
We have $e_k(\tau)=0$ whenever $k$ is odd.
For $k \ge 4$ the Eisenstein series $e_k(\tau)$ are modular forms of ${\mathcal M}_k(\mathrm{SL}_2({\mathbb Z}))$.
The space ${\mathcal M}_k(\mathrm{SL}_2({\mathbb Z}))$ has a basis of the form
\bq
\label{basis_SL2Z}
 \left( e_4\left(\tau\right) \right)^{\nu_4} \left(e_6\left(\tau\right)\right)^{\nu_6},
\eq
where $\nu_4$ and $\nu_6$ run over all non-negative integers with $4 \nu_4 + 6 \nu_6 = k$.

As an example, let us give the cusp form of modular weight $12$ for $\mathrm{SL}_2({\mathbb Z})$:
\bq
\label{def_Delta}
 \Delta\left(\tau\right)
 & = &
 \left(2\pi i\right)^{12} \eta\left(\tau\right)^{24}
 \;\; = \;\;
 10800 \left( 20 \left(e_4\left(\tau\right)\right)^3 - 49 \left(e_6\left(\tau\right)\right)^2 \right).
\eq
 
\subsubsection{Eisenstein series for $\Gamma_1(N)$}

Let $\Gamma$ be a congruence subgroup of $\mathrm{SL}_2({\mathbb Z})$.
By definition there exists an $N$, such that
\bq
 \Gamma\left(N\right) & \subseteq & \Gamma.
\eq
This implies
\bq
 {\mathcal M}_k\left(\Gamma\right) & \subseteq & {\mathcal M}_k\left(\Gamma\left(N\right)\right) 
\eq
and this reduces in a first step the study of modular forms for an arbitrary congruence subgroup $\Gamma$
to the study of modular forms of the principal congruence subgroup $\Gamma\left(N\right)$.
Now let $\eta(\tau) \in {\mathcal M}_k(\Gamma(N))$.
Then \cite{Miyake}
\bq
 \eta\left(N\tau\right)
 & \in &
 {\mathcal M}_k\left(\Gamma_1\left(N^2\right)\right),
\eq
which reduces in a second step the study of modular forms for an arbitrary congruence subgroup $\Gamma$
to the study of modular forms of the congruence subgroup $\Gamma_1\left(N\right)$.

Let us therefore consider modular forms for the congruence subgroups $\Gamma_1(N)$, and here
in particular the Eisenstein subspace $\mathcal{E}_k(\Gamma_1(N))$.
To this aim we define Eisenstein series with characters.
Of particular interest are characters which are obtained from the Kronecker symbol.
These characters take the values $\{-1,0,1\}$.
In general, the value of a Dirichlet character is a root of unity or zero.
The restriction to Dirichlet characters obtained from the Kronecker symbol
has the advantage that the $\bar{q}$-expansion of the Eisenstein series can be computed 
within the rational numbers.
The case of non-trivial roots of unity (e.g. of order three or higher) can be handled 
with the Eisenstein series defined in the next sub-section.

Let $a$ be an integer, which is either one or the discriminant of a quadratic field.
In appendix~\ref{appendix:Kronecker_symbol} we give a criteria for $a$ being the
discriminant of a quadratic field.
The Kronecker symbol, also defined in appendix~\ref{appendix:Kronecker_symbol}, then defines a primitive
Dirichlet character
\bq
 \chi_a\left(n\right) & = & 
 \left( \frac{a}{n} \right)
\eq
of conductor $|a|$.

Let $a$ and $b$ be as above (i.e. integers, which are either one or the discriminant of a quadratic field).
We set
\begin{align}
\label{def_Eisenstein_E}
E_{k,a,b}\left(\tau\right) &
 = 
 a_0 + \sum\limits_{n=1}^{\infty} \left( \sum\limits_{d|n} \chi_a(n/d) \cdot \chi_b(d) \cdot d^{k-1} \right) \bar{q}^{n},
\end{align}
The normalisation is such that the coefficient of $\bar{q}$ is one.
The constant term $a_0$ is given by
\begin{align}
 a_0 &
 = 
 \begin{cases}
 -\frac{B_{k,b}}{2k}, \qquad &\text{if}\ |a|=1, \\
0, \qquad &\text{if}\ |a|>1.
\end{cases}
\end{align}
Note that the constant term $a_0$ depends on $a$ and $b$.
The generalised Bernoulli numbers $B_{k,b}$ are defined by
\bq
\label{def_generalised_Bernoulli}
 \sum\limits_{n=1}^{|b|} \chi_b(n) \dfrac{xe^{nx}}{e^{|b|x}-1}
 & = &
 \sum\limits_{k=0}^{\infty} B_{k,b} \dfrac{x^k}{k!}. 
\eq
Note that in the case of the trivial character $\chi_1$,
eq.~(\ref{def_generalised_Bernoulli}) reduces to
\bq
 \dfrac{xe^{x}}{e^{x}-1}
 & = &
 \sum\limits_{k=0}^{\infty} B_{k,1} \dfrac{x^k}{k!},
\eq
yielding $B_{1,1}=1/2$.
The ordinary Bernoulli numbers $B_k$ are generated by $x/(e^x-1)$ 
(i.e. without an extra factor $e^x$ in the numerator)
and yield $B_1=-1/2$.

Let now $a$ and $b$ be such that 
\bq
\label{condition_k}
 \chi_a\left(-1\right) \chi_b\left(-1\right) & = & \left(-1\right)^k
\eq
and if $k=1$ one requires in addition
\bq
\label{condition_k_eq_1}
 \chi_a\left(-1\right) \; = \; 1,
 & &
 \chi_b\left(-1\right) \; = \; -1.
\eq
Let $N$ be an integer multiple of $K |a| |b|$.
We then set
\bq
\label{def_Eisenstein_E_B}
 E_{k,N,a,b,K}\left(\tau\right) 
 & = & 
 \left\{
  \begin{array}{lll}
    E_{k,a,b}\left(K\tau_N\right), & (k,a,b) \neq (2,1,1), & K \ge 1, \\ 
    E_{2,1,1}\left(\tau_N\right) - K E_{2,1,1}\left(K\tau_N\right), & (k,a,b) = (2,1,1), & K > 1. \\
  \end{array}
 \right.
\eq
$E_{k,N,a,b,K}(\tau)$ is a modular form for $\Gamma_1(N)$ of modular weight $k$ and level $N$:
\bq
 E_{k,N,a,b,K}\left(\tau\right)
 & \in & 
 \mathcal{E}_k(\Gamma_1(N)).
\eq
Remark: For $k$ even and $k \ge 4$ the relation between the Eisenstein series $e_k(\tau)$ defined in eq.~(\ref{def_e_k}) and the Eisenstein series with a trivial character is
\bq
\label{relation_e_k_to_E_k_1_1}
 e_k\left(\tau\right)
 & = &
 2 \frac{\left(2 \pi i \right)^k}{\left(k-1\right)!}
 E_{k,1,1}\left(\tau\right).
\eq

\subsubsection{Eisenstein series related to elliptic multiple polylogarithms}

Elliptic multiple polylogarithms, where all $z$-values are rational points, i.e. of the form
\bq
 z & = & r_1 + r_2 \tau,
 \;\;\;\;\;\;
 r_1, r_2\; \in \; {\mathbb Q},
 \;\;\;\;\;\;
 r_1 \; = \; \frac{r}{N},
 \;
 r_2 \; = \; \frac{s}{N},
\eq
are related to iterated integrals of Eisenstein series for the principal congruence subgroups $\Gamma(N)$ \cite{Broedel:2018iwv,Duhr:2019rrs}.
Of course, these Eisenstein series can always be written as a linear combination of a basis of the Eisenstein subspace $\mathcal{E}_k(\Gamma(N))$.
Nevertheless, it is convenient to have these Eisenstein series directly available.

Let $r, s$ be integers with $0 \le r,s <N$.
Following \cite{Broedel:2018iwv,Duhr:2019rrs} we set
\bq
\label{def_Eisenstein_h}
 h_{k,N,r,s}\left(\tau\right)
 & = &
 \sum\limits_{n=1}^\infty
 a_n
 \bar{q}^n_N.
\eq
For $n \ge 1$ the coefficients are given by
\bq
\label{coeff_h_a_n}
 a_n
 & = &
 \frac{1}{2 N^k}
 \sum\limits_{d|n}
 \sum\limits_{c_1=0}^{N-1}
 d^{k-1}
 \left[
   e^{\frac{2\pi i}{N} \left(r \frac{n}{d} - \left(s-d\right) c_1 \right)}
   + \left(-1\right)^k
   e^{-\frac{2\pi i}{N} \left(r \frac{n}{d} - \left(s+d\right) c_1 \right)}
 \right].
\eq
The constant term is given for $k\ge2$ by
\bq
 a_0
 & = & 
 - \frac{1}{2k} B_k\left(\frac{s}{N}\right),
\eq
where $B_k(x)$ is the $k$'th Bernoulli polynomial defined by
\bq
 \frac{t e^{x t}}{e^t-1}
 & = &
 \sum\limits_{k=0}^\infty \frac{B_k\left(x\right)}{k!} t^k.
\eq
For $k=1$ the constant term is given by
\bq
 a_0
 & = &
 \left\{
 \begin{array}{ll}
  \frac{1}{4} - \frac{s}{2N}, & s \neq 0, \\
 0, & (r,s) = (0,0), \\
 \frac{i}{4} \cot\left(\frac{r}{N} \pi \right),
 & \mbox{otherwise}.
 \end{array}
 \right.
\eq
The normalisation used here is compatible with the normalisation of the previous subsection.
For example we have
\bq
 h_{k,1,0,0}\left(\tau\right)
 & = &
 E_{k,1,1}\left(\tau\right).
\eq
The normalisation differs by a factor $-(k-1)!/2/(2\pi i)^k$ from the one used in ref.~\cite{Broedel:2018iwv,Duhr:2019rrs}, i.e.
\bq
 h_{k,N,r,s}\left(\tau\right)
 & = &
 - \frac{1}{2} \frac{\left(k-1\right)!}{\left(2 \pi i \right)^k} h^{(k)}_{N,r,s}\left(\tau\right),
\eq
where $h^{(k)}_{N,r,s}$ denotes the quantity defined in ref.~\cite{Broedel:2018iwv,Duhr:2019rrs}.

With the exception of $(k,r,s) \neq (2,0,0)$ the $h_{k,N,r,s}(\tau)$ are Eisenstein series for $\Gamma(N)$:
\bq
 h_{k,N,r,s}\left(\tau\right) 
 & \in & 
 \mathcal{E}_k\left(\Gamma\left(N\right)\right).
\eq
For $(k,N,r,s)=(2,1,0,0)$ we have
\bq
 h_{2,1,0,0}\left(\tau\right)
 & = &
 E_{2,1,1}\left(\tau\right)
 \; = \;
 \frac{1}{2\left(2\pi i\right)^2} e_2\left(\tau\right),
\eq
which is not a modular form.

Please note that while the coefficients of the $\bar{q}$-expansion in eq.~(\ref{def_Eisenstein_E}) are rational numbers
and can be computed entirely within the rational numbers, the coefficients
of the $\bar{q}$-expansion in eq.~(\ref{coeff_h_a_n}) involve roots of unity.

On the positive side, the Eisenstein series $h_{k,N,r,s}(\tau)$ have a simple
transformation law under the full modular group $\mathrm{SL}_2({\mathbb Z})$.
For $\gamma \in \mathrm{SL}_2\left({\mathbb Z}\right)$ we have
\bq
 h_{k,N,r,s}\left(\frac{a\tau+b}{c\tau+d}\right)
 & = &
 \left(c\tau +d\right)^k
 h_{k,N,rd+sb \bmod N,rc+sa \bmod N}\left(\tau\right),
 \nonumber \\
 & &
 \gamma \; = \;
 \left(\begin{array}{cc} a & b \\ c & d \\ \end{array} \right)
 \; \in \;
 \mathrm{SL}_2\left({\mathbb Z}\right).
\eq


\section{Integration kernels}
\label{sect:kernels}

With the notation for the special functions at hand we now define various integration kernels.

\subsection{Multiple polylogarithms}

A prominent example of iterated integrals are multiple polylogarithms, 
where 
$M={\mathbb C}$ with coordinate $y$ and
\bq
\label{def_omega_mpl}
 \omega^{\mathrm{mpl}}\left(z_j\right) & = & \frac{dy}{y-z_j}.
\eq
The $z_j$'s are complex parameters, only subject to the restriction that there are no poles along the 
integration path.
The standard notation for these specific iterated integrals is
\bq
 G\left(z_1,\dots,z_r;y\right)
 & = &
 I\left(\omega^{\mathrm{mpl}}\left(z_1\right),...,\omega^{\mathrm{mpl}}\left(z_r\right);y\right).
\eq
The integrands have a convergent series expansion for
\bq
 \left| y \right| & < & \left| z_j \right|.
\eq

\subsection{Modular forms}

Iterated integrals of modular forms are a second example.
These type of iterated integrals occur in the equal-mass sunrise integral \cite{Adams:2017ejb}. 
Let $\eta_k(\tau)$ be a modular form of modular weight $k$.
We now take $M={\mathbb H}$ 
and consider the integration path $\gamma$ from $\tau_i=i \infty$ to $\tau_f=\tau$.
Under the change of variables $\tau \rightarrow \bar{q}$ (given by eq.~(\ref{def_basic_variables}))
we integrate from $\bar{q}_i=0$ to $\bar{q}_f=\bar{q}$.
For modular forms of level $N$ we set $\tau_N=\tau/N$ and 
\bq
 \bar{q}_N
 & = &
 e^{2\pi i \tau_N}
 \;\; = \;\;
 e^{\frac{2\pi i \tau}{N}}.
\eq
For a generic modular form $\eta_k$ of modular weight $k$ and level $N$ we set
\bq
 \omega^{\mathrm{modular}}\left(\eta_k\right)
 & = &
 2 \pi i \; C_k \; \eta_k\left(\tau\right) d\tau_N
 \;\; = \;\;
 C_k \; \eta_k\left(\tau\right) \frac{d\bar{q}_N}{\bar{q}_N},
\eq
where $C_k$ is a constant defining the normalisation.
The default value will be $C_k=1$.
If the modular form $\eta_k(\tau)$ has the $\bar{q}_N$-expansion
\bq
 \eta_k\left(\tau\right)
 & = &
 \sum\limits_{n=0}^\infty a_n \bar{q}_N^n,
\eq
we have
\bq
 \omega^{\mathrm{modular}}\left(\eta_k\right)
 & = &
 C_k \; \sum\limits_{n=0}^\infty a_n \bar{q}_N^{n-1} d\bar{q}_N.
\eq
$\omega^{\mathrm{modular}}(\eta_k)$ has a trailing zero, if $\eta_k(\tau)$ does not vanish at the cusp $\tau=i \infty$.

We are mainly interested in the case where $\eta_k(\tau)$ is an Eisenstein series with characters $E_{k,N,a,b,K}(\tau)$ as defined in eq.~(\ref{def_Eisenstein_E_B}).
We set
\bq
\label{def_omega_Eisenstein}
 \omega^{\mathrm{Eisenstein}}_{k,N,a,b,K} & = & 
 2\pi i \; C_k \; E_{k,N,a,b,K}\left(\tau\right) d\tau_N
 \;\; = \;\;
 C_k \; E_{k,N,a,b,K}\left(\tau\right) \frac{d\bar{q}_N}{\bar{q}_N}.
\eq
For the Eisenstein series defined in eq.~(\ref{def_Eisenstein_h}) we set
\bq
\label{def_omega_Eisenstein_h}
 \omega^{\mathrm{Eisenstein,h}}_{k,N,r,s} & = & 
 2\pi i \; C_k \; h_{k,N,r,s}\left(\tau\right) d\tau_N
 \;\; = \;\;
 C_k \; h_{k,N,r,s}\left(\tau\right) \frac{d\bar{q}_N}{\bar{q}_N}.
\eq
The integrands of $\omega^{\mathrm{Eisenstein}}_{k,N,a,b,K}$ and $\omega^{\mathrm{Eisenstein,h}}_{k,N,r,s}$ have a convergent series expansion for
\bq
 \left| \bar{q}_N \right| & < & 1.
\eq
If $\eta^{(1)}_{k_1}$ is a modular form of modular weight $k_1$
and $\eta^{(2)}_{k_2}$ is a modular form of modular weight $k_2$,
then the product
\bq
 \eta^{(1)}_{k_1} \eta^{(2)}_{k_2}
\eq
is a modular form of weight $k_1+k_2$.
More generally, 
let $P_k(\eta^{(1)}_{k_1}, \dots, \eta^{(r)}_{k_r})$ be a polynomial in the modular forms
$\eta^{(1)}_{k_1}, \dots, \eta^{(r)}_{k_r}$, all of level $N$, 
such that each term in the expanded polynomial has the same modular weight $k$.
We set
\bq
\label{def_omega_modular}
 \omega^{\mathrm{modular}}\left(P_k\left(\eta^{(1)}_{k_1}, \dots, \eta^{(r)}_{k_r}\right)\right)
 & = &
 2 \pi i \; C_k \; P_k\left(\eta^{(1)}_{k_1}, \dots, \eta^{(r)}_{k_r}\right) d\tau_N
 \nonumber \\
 & = &
 C_k \; P_k\left(\eta^{(1)}_{k_1}, \dots, \eta^{(r)}_{k_r}\right) \frac{d\bar{q}_N}{\bar{q}_N}.
\eq
The convergence properties are inherited from above:
\bq
 \left| \bar{q}_N \right| & < & 1.
\eq
Allowing for polynomials in modular forms is useful: For example we have already seen
in eq.~(\ref{basis_SL2Z}) that the space of modular
forms for $\mathrm{SL}_2({\mathbb Z})$ is generated by monomials in $e_4(\tau)$ and $e_6(\tau)$.

\subsection{Elliptic multiple polylogarithms}

Let us now generalise multiple polylogarithms to the elliptic setting.
This generalisation is commonly named elliptic multiple polylogarithms, 
but as the definitions used by various authors differ in the details,
we carefully explain the variant relevant to Feynman integrals.

Let us start from multiple polylogarithms.
We may view multiple polylogarithms as iterated integrals on a covering space of ${\mathcal M}_{0,n}$,
the latter being the moduli space of a Riemann sphere with $n$ marked points.
In the same spirit we view elliptic multiple polylogarithms as iterated integrals 
on a covering space of ${\mathcal M}_{1,n}$,
the latter being the moduli space of a Riemann surface of genus one with $n$ marked points.

It is not possible that the integration kernels are double-periodic and meromorphic at the same time
and we can only require one of these two properties.
This is the main point, where the various available definitions in the literature differ:
Different authors require either double-periodicity or meromorphicity.
For the application towards Feynman integrals we want meromorphicity.
The integration kernels are then either multi-valued functions on an elliptic curve or single-valued function
on a covering space.

In the genus zero case the dimension of the moduli space ${\mathcal M}_{0,n}$
is
\bq
 \dim {\mathcal M}_{0,n} & = & n-3.
\eq
If $z_0,z_1,\dots,z_{n-1}$ are the marked points on the Riemann sphere, we may use M\"obius transformations
to fix $z_0=0$, $z_{n-2}=1$ and $z_{n-1}=\infty$. Thus
\bq
 \left( z_1, z_2, \dots, z_{n-3} \right)
\eq
are standard coordinates on ${\mathcal M}_{0,n}$.
The dimension of the 
moduli space ${\mathcal M}_{1,n}$
is
\bq
 \dim {\mathcal M}_{1,n} & = & n.
\eq
If $z_0,z_1,\dots,z_{n-1}$ are the marked points on the Riemann surface of genus one, we may use 
translation symmetry
to fix $z_0=0$.
This gives us 
\bq
 \left( z_1, z_2, \dots, z_{n-1},\tau \right)
\eq
as standard coordinates on ${\mathcal M}_{1,n}$.

In principle we may consider iterated integrals along arbitrary integration paths.
However, in practice we only consider a few standard integration paths.
An arbitrary integration path on a covering space of ${\mathcal M}_{1,n}$ can be decomposed
into pieces along the $\tau$-direction and pieces along the $z_j$-directions.
The latter we always pull-back to a one-dimensional space.
Thus we consider the integration path to be either along $\tau$ or along a $z$-variable.
From the differential equation for the unequal mass sunrise integral we know 
that the differential one-forms enter in the linear
combination
\bq
\label{def_omega}
 \omega_k\left(z, K \tau\right)
 & = &
 C_k \left(2\pi i\right)^{2-k}
 \left[
  g^{(k-1)}\left(z, K \tau\right) dz + K \left(k-1\right) g^{(k)}\left(z, K \tau\right) \frac{d\tau}{2\pi i}
 \right].
\eq
The functions $g^{(k)}(z,\tau)$ have been defined in eq.~(\ref{def_g_n}).

\subsubsection{Integration along $\tau$}

For the integration along $\tau$ we take the part proportional to $d\tau$ in eq.~(\ref{def_omega}).
We set
\bq
\label{def_omega_Kronecker_tau}
 \omega^{\mathrm{Kronecker},\tau}_{k,K}\left(z_j\right)
 & = &
 C_k \left(2\pi i\right)^{2-k} K \left(k-1\right) g^{(k)}\left(z_j,K \tau\right) \frac{d\tau}{2\pi i}
 \nonumber \\
 & = &
 \frac{C_k K \left(k-1\right)}{\left(2\pi i\right)^{k}} g^{(k)}\left(z_j,K \tau\right) \frac{d\bar{q}}{\bar{q}}.
\eq
The $\bar{q}$-expansion follows from eq.~(\ref{g_n_explicit}).
Explicitly we have with $\bar{w}_j=\exp(2\pi i z_j)$
\bq
\label{q_expansion_omega}
 \omega^{\mathrm{Kronecker},\tau}_{0,K}\left(z_j\right)
 & = &
 - C_0 K \frac{d\bar{q}}{\bar{q}},
 \nonumber \\
 \omega^{\mathrm{Kronecker},\tau}_{1,K}\left(z_j\right)
 & = &
 0,
 \nonumber \\
 \omega^{\mathrm{Kronecker},\tau}_{k,K}\left(z_j\right)
 & = &
 - \frac{C_k K}{\left(k-2\right)!} 
 \left[
 - \frac{B_k}{k}
       + \sum\limits_{n=1}^\infty c_n^{(k)}\left(\bar{w}_j\right) \cdot \bar{q}^{K n}
 \right] \frac{d\bar{q}}{\bar{q}},
 \;\;\;\;\;\;\;\;\;\;\;\;\;\;\;
 k > 1,
\eq
with
\bq
 c_n^{(k)}\left(\bar{w}\right)
 & = &
 \sum\limits_{j|n}
 \left[ \bar{w}^j + \left(-1\right)^k \bar{w}^{-j} \right]
 \left( \frac{n}{j} \right)^{k-1}.
\eq
The integrand of $\omega^{\mathrm{Kronecker},\tau}_{0,K}(z_j)$ has a convergent series expansion for
\bq
 \left| \bar{q} \right| \; < \; 1,
 & &
 - \mathrm{Im}\left(K\tau\right) \; < \; \mathrm{Im}\left(z_j\right) \; < \; \mathrm{Im}\left(K\tau\right).
\eq

\subsubsection{Integration along $z$}

For the integration along $z$ we take the part proportional to $dz$ in eq.~(\ref{def_omega}).
In addition, we allow for a translation in $z$.
We set 
\bq
\label{def_omega_Kronecker_z}
 \omega^{\mathrm{Kronecker},z}_{k,K}\left(z_j,\tau\right)
 & = &
 C_k \left(2\pi i\right)^{2-k}
 g^{(k-1)}\left(z-z_j, K \tau\right) dz.
\eq
For the corresponding iterated integrals we need
the Laurent expansion in $z$.
The differential form $\omega^{\mathrm{Kronecker},z}_{0,K}(z_j,\tau)$ vanishes. 
For $k=1$ we have
\bq
 \omega^{\mathrm{Kronecker},z}_{1,K}\left(z_j,\tau\right)
 & = &
 2 \pi i C_1 dz.
\eq
For $k=2$, $z_j \neq 0$ we have with $\bar{w}_j=\exp(2\pi i z_j)$
\bq
\lefteqn{
 \omega^{\mathrm{Kronecker},z}_{2,K}\left(z_j,\tau\right)
 = } & &
 \nonumber \\
 & &
 - 2 \pi i C_2
 \sum\limits_{n=0}^\infty \frac{\left(2\pi i\right)^n}{n!} 
         \left[ \frac{1}{2} \mathrm{Li}_{-n}\left(\bar{w}_j^{-1}\right) - \frac{\left(-1\right)^n}{2}\mathrm{Li}_{-n}\left(\bar{w}_j\right) + \overline{\mathrm{E}}_{-n;0}\left(\bar{w}_j^{-1};1;\bar{q}^K\right) \right] z^n
 dz.
\eq
For $n\in{\mathbb N}_0$ the function $\mathrm{Li}_{-n}(x)$ is a rational function in $x$ and given by
\bq
 \mathrm{Li}_{-n}\left(x\right)
 & = &
 \left( x \frac{d}{dx} \right)^n \frac{x}{1-x}.
\eq
Furthermore (again for $n\in{\mathbb N}_0$) 
\bq
 \mathrm{Li}_{-n}\left(x^{-1}\right)
 & = &
 - \delta_{n,0}
 - \left(-1\right)^{n} \mathrm{Li}_{-n}\left(x\right).
\eq
For $k=2$, $z_j = 0$ we have
\bq
\lefteqn{
 \omega^{\mathrm{Kronecker},z}_{2,K}\left(0,\tau\right)
 = } & &
 \nonumber \\
 & &
 - 2 \pi i C_2 \left\{
   -\frac{1}{2\pi i} \frac{1}{z}
   + \sum\limits_{n=1}^\infty \frac{\left(2\pi i\right)^n}{n!} 
         \left[ - \frac{B_{n+1}}{n+1} + \overline{\mathrm{E}}_{-n;0}\left(1;1;\bar{q}^K\right) \right] z^n
  \right\} dz.
\eq
For $k > 2$ we have
\bq
 \omega^{\mathrm{Kronecker},z}_{k,K}\left(z_j,\tau\right)
 & = &
 - \frac{2 \pi i C_k}{\left(k-2\right)!} 
 \left\{
   - \frac{B_{k-1}}{k-1} 
   + \sum\limits_{n=0}^\infty \frac{\left(2\pi i\right)^n}{n!} 
         \overline{\mathrm{E}}_{-n;2-k}\left(\bar{w}_j^{-1};1;\bar{q}^K\right) z^n
 \right\} dz.
\eq
The integrand of $\omega^{\mathrm{Kronecker},z}_{k,K}\left(z_j,\tau\right)$ has a convergent series expansion for
\bq
 \left| \bar{q} \right| \; < \; 1,
 \;\;\;\;\;\;
 \left| z \right| \; < \; \left| z_j \right|,
 \;\;\;\;\;\;
 - \mathrm{Im}\left(K\tau\right) \; < \; \mathrm{Im}\left(z_j\right) \; < \; \mathrm{Im}\left(K\tau\right),
 & &
 \nonumber \\
 - \mathrm{Im}\left(K\tau\right) \; < \; \mathrm{Im}\left(z-z_j\right) \; < \; \mathrm{Im}\left(K\tau\right).
\eq
Ref.~\cite{Broedel:2017kkb} defines elliptic multiple polylogarithms 
$\widetilde{\Gamma}\!\left({\begin{smallmatrix} n_1 & ... & n_r \\ z_1 & ... & z_r \\ \end{smallmatrix}}; z; \tau \right)$
as iterated integrals on an elliptic curve
with fixed modular parameter $\tau$ recursively through
\bq
\label{def_Gammatilde}
 \widetilde{\Gamma}\!\left({\begin{smallmatrix} n_1 & ... & n_r \\ z_1 & ... & z_r \\ \end{smallmatrix}}; z; \tau \right)
 & = &
 \int\limits_0^z dz' \; g^{(n_1)}(z'-z_1, \tau) \;
 \widetilde{\Gamma}\!\left({\begin{smallmatrix} n_2 & ... & n_r \\ z_2 & ... & z_r \\ \end{smallmatrix}}; z'; \tau \right),
 \nonumber \\
 \widetilde{\Gamma}\!\left(; z; \tau \right)
 & = &
 1.
\eq
For the default choice of the normalisation factors $C_k=1$
we have the relation
\bq
\label{elliptic_polylogs_I}
 \widetilde{\Gamma}\!\left({\begin{smallmatrix} n_1 & ... & n_r \\ z_1 & ... & z_r \\ \end{smallmatrix}}; z; \tau \right)
 & = &
 \left( 2 \pi i\right)^{n_1+\dots+n_r-r}
 I\left( \omega^{\mathrm{Kronecker},z}_{n_1+1,1}\left(z_1,\tau\right), \dots, \omega^{\mathrm{Kronecker},z}_{n_r+1,1}\left(z_r,\tau\right); z \right),
\eq
while for $C_k=(2\pi i)^{k-2}$ we have
\bq
\label{elliptic_polylogs_II}
 \widetilde{\Gamma}\!\left({\begin{smallmatrix} n_1 & ... & n_r \\ z_1 & ... & z_r \\ \end{smallmatrix}}; z; \tau \right)
 & = &
 I\left( \omega^{\mathrm{Kronecker},z}_{n_1+1,1}\left(z_1,\tau\right), \dots, \omega^{\mathrm{Kronecker},z}_{n_r+1,1}\left(z_r,\tau\right); z \right).
\eq
Please note the shift of indices $(n_j+1)$ in $\omega^{\mathrm{Kronecker},z}_{n_j+1,1}(z_j,\tau)$
on the right-hand side of eq.~(\ref{elliptic_polylogs_I}) and eq.~(\ref{elliptic_polylogs_II}).

\subsection{$\mathrm{ELi}$-kernel}

The functions 
\bq
 \mathrm{ELi}_{n;m}\left(\bar{u};\bar{v};\bar{q}\right) & = & 
 \sum\limits_{j=1}^\infty \sum\limits_{k=1}^\infty \; \frac{\bar{u}^j}{j^n} \frac{\bar{v}^k}{k^m} \bar{q}^{j k}.
\eq
were introduced in \cite{Adams:2014vja} as a generalisation of the classical polylogarithms 
$\mathrm{Li}_{n}(\bar{u})$.
The function $\mathrm{ELi}_{n;m}(\bar{u};\bar{v};\bar{q})$
depends on three variables $\bar{u}$, $\bar{v}$, $\bar{q}$ and two (integer) indices $n$, $m$ and
is symmetric under the exchange of the pair $(\bar{u},n)$ with $(\bar{v},m)$.
The two summations are coupled through the variable $\bar{q}$.
We define integration kernels associated to these functions by
\bq
\label{def_omega_ELi}
 \omega^{\mathrm{ELi}}_{n;m}\left(\bar{u};\bar{v}\right)
 & = &
 \mathrm{ELi}_{n;m}\left(\bar{u};\bar{v};\bar{q}\right) \frac{d\bar{q}}{\bar{q}}
 \;\; = \;\;
 \sum\limits_{j=1}^\infty \sum\limits_{k=1}^\infty \; \frac{\bar{u}^j}{j^n} \frac{\bar{v}^k}{k^m} \bar{q}^{j k - 1} d\bar{q}.
\eq
The integrand of $\omega^{\mathrm{ELi}}_{n;m}\left(\bar{u};\bar{v}\right)$ has a convergent series expansion for
\bq
 \left| \bar{q} \right| \; < \; 1,
 & &
 \left|\bar{u}\right|,
 \left|\bar{v}\right|
 \; < \; \left|\bar{q}\right|^{-1}.
\eq
In \cite{Adams:2015ydq} we introduced a multi-variable generalisation
\bq
\lefteqn{
 \mathrm{ELi}_{n_1,...,n_l;m_1,...,m_l;2o_1,...,2o_{l-1}}\left(\bar{u}_1,...,\bar{u}_l;\bar{v}_1,...,\bar{v}_l;q\right) 
 = }
 & & \nonumber \\
 & = &
 \sum\limits_{j_1=1}^\infty ... \sum\limits_{j_l=1}^\infty
 \sum\limits_{k_1=1}^\infty ... \sum\limits_{k_l=1}^\infty
 \;\;
 \frac{\bar{u}_1^{j_1}}{j_1^{n_1}} ... \frac{\bar{u}_l^{j_l}}{j_l^{n_l}}
 \;\;
 \frac{\bar{v}_1^{k_1}}{k_1^{m_1}} ... \frac{\bar{v}_l^{k_l}}{k_l^{m_l}}
 \;\;
 \frac{q^{j_1 k_1 + ... + j_l k_l}}{\prod\limits_{i=1}^{l-1} \left(j_i k_i + ... + j_l k_l \right)^{o_i}}.
\eq
This multiple sum can be written as an iterated integral with the integration kernels defined above
as
\bq
\lefteqn{
 \mathrm{ELi}_{n_1,...,n_l;m_1,...,m_l;2o_1,...,2o_{l-1}}\left(\bar{u}_1,...,\bar{u}_l;\bar{v}_1,...,\bar{v}_l;q\right) 
 = }
 & & \nonumber \\
 & = &
 I_{o_1,\dots,o_{l-1},1}\left( \omega^{\mathrm{ELi}}_{n_1;m_1}\left(\bar{u}_1;\bar{v}_1\right), \dots, \omega^{\mathrm{ELi}}_{n_{l-1};m_{l-1}}\left(\bar{u}_{l-1};\bar{v}_{l-1}\right), \omega^{\mathrm{ELi}}_{n_l-1;m_l-1}\left(\bar{u}_l;\bar{v}_l\right); \bar{q} \right).
\eq

\subsection{$\overline{\mathrm{E}}$-kernel}

The functions
\bq
 \overline{\mathrm{E}}_{n;m}\left(\bar{u};\bar{v};q\right) 
 & = &
  \mathrm{ELi}_{n;m}\left(\bar{u};\bar{v};q\right)
  - \left(-1\right)^{n+m} \mathrm{ELi}_{n;m}\left(\bar{u}^{-1};\bar{v}^{-1};q\right)
\eq
are linear combinations of the functions $\mathrm{ELi}_{n;m}(\bar{u};\bar{v};\bar{q})$.
As this particular linear combination occurs frequently, 
and in particular 
in the $\bar{q}$-expansion of the functions $g^{(n)}(z,\tau)$,
it is advantageous to define 
the integration kernels
\bq
\label{def_omega_Ebar}
 \omega^{\overline{\mathrm{E}}}_{n;m}\left(\bar{u};\bar{v}\right)
 & = &
 \overline{\mathrm{E}}_{n;m}\left(\bar{u};\bar{v};\bar{q}\right) \frac{d\bar{q}}{\bar{q}}
 \;\; = \;\;
 \sum\limits_{j=1}^\infty \sum\limits_{k=1}^\infty \; \frac{1}{j^n k^m} \left[ \bar{u}^j \bar{v}^k - \left(-1\right)^{n+m} \bar{u}^{-j} \bar{v}^{-k} \right] \bar{q}^{j k - 1} d\bar{q}.
\eq
The integrand of $\omega^{\overline{\mathrm{E}}}_{n;m}\left(\bar{u};\bar{v}\right)$ has a convergent series expansion for
\bq
 \left| \bar{q} \right| \; < \; 1,
 & &
 \left|\bar{q}\right| \; < \; 
 \left|\bar{u}\right|,
 \left|\bar{v}\right|
 \; < \; \left|\bar{q}\right|^{-1}.
\eq

\subsection{User-defined kernel}

Our last example is a user-defined kernel.
Let $f(y)$ be a function, which has a Laurent expansion around $y=0$ and at $y=0$ maximally a simple
pole.
We then set
\bq
\label{def_omega_user}
 \omega^{\mathrm{user}}\left(f\right)
 & = & 
 f\left(y\right) dy.
\eq
An example is
\bq
 f\left(y\right)
 & = &
 \frac{1}{\sqrt{\left(y-z_1\right)\left(y-z_2\right)\left(y-z_3\right)\left(y-z_4\right)}}.
\eq
A typical application is an iterated integral, where only the outermost integration involves
a user-defined kernel, while all other integration kernels are of multiple polylogarithm-type:
\bq
 I\left(\omega^{\mathrm{user}}\left(f\right),\omega^{\mathrm{mpl}}\left(z_2\right),...,\omega^{\mathrm{mpl}}\left(z_r\right);y\right).
\eq
These type of iterated integrals have been considered for example in \cite{Remiddi:2017har,Bourjaily:2017bsb}.

\section{Implementation}
\label{sect:implementation}

The numerical evaluations have been implemented as part of GiNaC \cite{Bauer:2000cp}, a C++
library for computer algebra ({\tt http://www.ginac.de}). 
The GiNaC library is open source software and freely available.
GiNaC enables symbolic algebraic manipulations within the C++ programming language.
Like FORM \cite{Vermaseren:2000nd}, it was developed within the high-energy physics community.
GiNaC allows numerics in arbitrary precision.

\subsection{Complete elliptic integrals}

In order to compute the periods and pseudo-periods of an elliptic curve one needs the complete elliptic integrals of the first and second kind.
The complete elliptic integral of the first kind is denoted $K(k)$, the complete elliptic integral of the second kind is denoted $E(k)$.
The definition is given in appendix~\ref{appendix:standard_functions}.
In GiNaC these functions are called
\begin{verbatim}
 EllipticK(k);
 EllipticE(k);
\end{verbatim}
The numerical evaluation is based on the arithmetic-geometric mean.
The method is described in appendix~\ref{appendix:agm}.

\subsection{Iterated integrals}

The interface for the iterated integrals follows closely the one for multiple polylogarithms.
We recall that for {\tt z1}, {\tt z2} and {\tt y} of type {\tt numeric}, the multiple polylogarithm
$G(z_1,z_2;y)$ is evaluated in GiNaC by
\begin{verbatim}
 G(lst{z1,z2},y);
\end{verbatim}
Now let {\tt omega1} and {\tt omega2} be of the type \verb|integration_kernel| or a subclass thereof
and {\tt y} as above.
The iterated integral $I(\omega_1,\omega_2;y)$ is evaluated in GiNaC by
\begin{verbatim}
 iterated_integral(lst{omega1,omega2},y);
\end{verbatim}
or by
\begin{verbatim}
 iterated_integral(lst{omega1,omega2},y,N_trunc);
\end{verbatim}
In the first case the truncation criterion of eq.~(\ref{truncation_criterion}) is used, in the second case the series is truncated at \verb|N_trunc|.
The available integration kernels are:
\begin{enumerate}
\item
\begin{verbatim}
integration_kernel();
\end{verbatim}
This is the base class and corresponds to the integration kernel $\omega=dz$.
\item
\begin{verbatim}
basic_log_kernel();
\end{verbatim}
This implements the integration kernel 
\bq
 L_0 & = & \frac{dz}{z}.
\eq
Please note that the integration variable is a dummy variable, which does not need to be specified.
Therefore, this class also represents $d\bar{q}/\bar{q}$.
\item
\begin{verbatim}
multiple_polylog_kernel(z_j);
\end{verbatim}
This defines the integration kernel $\omega^{\mathrm{mpl}}\left(z_j\right)$ as in eq.~(\ref{def_omega_mpl}).
\item
\begin{verbatim}
ELi_kernel(n, m, ubar, vbar);
\end{verbatim}
This defines the integration kernel $\omega^{\mathrm{ELi}}_{n;m}\left(\bar{u};\bar{v}\right)$ as in eq.~(\ref{def_omega_ELi}).
\item
\begin{verbatim}
Ebar_kernel(n, m, ubar, vbar);
\end{verbatim}
This defines the integration kernel $\omega^{\overline{\mathrm{E}}}_{n;m}\left(\bar{u};\bar{v}\right)$ as in eq.~(\ref{def_omega_Ebar}).
\item
\begin{verbatim}
Kronecker_dtau_kernel(k, z_j, K, C_k);
\end{verbatim}
This defines the integration kernel $\omega^{\mathrm{Kronecker},\tau}_{k,K}\left(z_j\right)$ as in eq.~(\ref{def_omega_Kronecker_tau}).
The last variable or the last two variables can be omitted.
The default values are $K=1$ and $C_k=1$.
\item
\begin{verbatim}
Kronecker_dz_kernel(k, z_j, tau, K, C_k);
\end{verbatim}
This defines the integration kernel $\omega^{\mathrm{Kronecker},z}_k\left(z_j,\tau\right)$ as in eq.~(\ref{def_omega_Kronecker_z}).
The last variable or the last two variables can be omitted.
The default values are $K=1$ and $C_k=1$.
\item
\begin{verbatim}
Eisenstein_kernel(k, N, a, b, K, C_k);
\end{verbatim}
This defines the integration kernel $\omega^{\mathrm{Eisenstein}}_{k,N,a,b,K}$ as in eq.~(\ref{def_omega_Eisenstein}).
The normalisation constant $C_k$ can be omitted. The default value is $C_k=1$.
The method 
\begin{verbatim}
 q_expansion_modular_form(qbar_N, order)
\end{verbatim}
gives the $\bar{q}_N$-expansion of $E_{k,N,a,b,K}(\tau)$ to order {\tt order}.
\verb|qbar_N| is a {\tt symbol}.
\item
\begin{verbatim}
Eisenstein_h_kernel(k, N, r, s, C_k);
\end{verbatim}
This defines the integration kernel $\omega^{\mathrm{Eisenstein,h}}_{k,N,r,s}$ as in eq.~(\ref{def_omega_Eisenstein_h}).
The normalisation constant $C_k$ can be omitted. The default value is $C_k=1$.
The method 
\begin{verbatim}
 q_expansion_modular_form(qbar_N, order)
\end{verbatim}
gives the $\bar{q}_N$-expansion of $h_{k,N,r,s}(\tau)$ to order {\tt order}.
\verb|qbar_N| is a {\tt symbol}.
\item
\begin{verbatim}
modular_form_kernel(k, P, C_k);
\end{verbatim}
This defines the integration kernel $\omega^{\mathrm{modular}}\left(P_k\right)$ as in eq.~(\ref{def_omega_modular}).
The normalisation constant $C_k$ can be omitted. The default value is $C_k=1$.
The method 
\begin{verbatim}
 q_expansion_modular_form(qbar_N, order)
\end{verbatim}
gives the $\bar{q}_N$-expansion of $P_k(\eta^{(1)}_{k_1}, \dots, \eta^{(r)}_{k_r})$ to order {\tt order}.
\verb|qbar_N| is a {\tt symbol}.
\item
\begin{verbatim}
user_defined_kernel(f, y);
\end{verbatim}
This defines the integration kernel $\omega^{\mathrm{user}}\left(f\right)$ as in eq.~(\ref{def_omega_user}).
{\tt y} is a {\tt symbol}.
\end{enumerate}

\section{Examples}
\label{sect:examples}

In this section we give several examples on how to use the program.

\subsection{Evaluating iterated integrals}

\subsubsection{Example 1}

We start with an example of an iterated integral of modular forms.
Let
\bq
 \omega_0 & = & \frac{d\bar{q}_6}{\bar{q}_6},
 \nonumber \\
 \omega^{\mathrm{modular}}_3 & = & \left[ E_{3,6,-3,1,1}\left(\tau\right) - 8 E_{3,6,-3,1,2}\left(\tau\right) \right] \frac{d\bar{q}_6}{\bar{q}_6}.
\eq
We let $\tau_6=\tau/6$. Then $\bar{q}_6=\exp(2\pi i \tau/6)=\exp(2\pi i \tau_6)$.
$\omega^{\mathrm{modular}}_3$ is a modular form of weight $3$ for $\Gamma_1(6)$.
Suppose we would like to evaluate the iterated integral
\bq
\label{example_1}
 I\left(\omega_0,\omega^{\mathrm{modular}}_3;\bar{q}_6\right).
\eq
This iterated integral is not unrelated to physics, 
it occurs in the equal-mass sunrise integral in two space-time dimensions.
The following program shows how to evaluate this integral numerically:
\begin{verbatim}
#include <ginac/ginac.h>

int main()
{
  using namespace std;
  using namespace GiNaC;

  Digits = 30;

  ex tau_6 = I;
  ex qbar_6 = exp(2*Pi*I*tau_6);
    
  ex omega_0 = basic_log_kernel();

  ex eta_1   = Eisenstein_kernel(3, 6, -3, 1, 1);
  ex eta_2   = Eisenstein_kernel(3, 6, -3, 1, 2);
  ex omega_3 = modular_form_kernel(3, eta_1-8*eta_2);

  ex expr = iterated_integral(lst{omega_0,omega_3},qbar_6);

  cout << "I = " << expr.evalf() << endl;

  return 0;
}
\end{verbatim}
Running this program will print out
\begin{verbatim}
I = 0.001863090057835543048808657035227425650174
\end{verbatim}
GiNaC offers a simple interactive shell \verb|ginsh|.
The same can be done in \verb|ginsh|:
\begin{verbatim}
> Digits=30;
30
> q6=evalf(exp(-2*Pi));
0.00186744273170798881443021293482703039343
> omega_0=basic_log_kernel(void);
basic_log_kernel()
> eta_1=Eisenstein_kernel(3,6,-3,1,1);
Eisenstein_kernel(3,6,-3,1,1,1)
> eta_2=Eisenstein_kernel(3,6,-3,1,2);
Eisenstein_kernel(3,6,-3,1,2,1)
> omega_3=modular_form_kernel(3,eta_1-8*eta_2,1);
modular_form_kernel(3,-8*Eisenstein_kernel(3,6,-3,1,2,1)
                      +Eisenstein_kernel(3,6,-3,1,1,1),1)
> iterated_integral({omega_0,omega_3},q6);
0.001863090057835543048808657035227425650174
\end{verbatim}

\subsubsection{Example 2}

As a second example we consider an iterated integral involving the Kronecker function $g^{(k)}(z_j,\tau)$.
With $\omega_0=d\bar{q}_6/\bar{q}_6$ as above and $C_3=i/\sqrt{3}$ 
we consider
\bq
 \omega^{\mathrm{Kronecker}}_3 
 & = & 
 \frac{i}{\sqrt{3} \left(2\pi i\right)^3}
 \left[ 2 g^{(3)}\left(\frac{1}{3},\tau_6\right) - 16 g^{(3)}\left(\frac{1}{3},2\tau_6\right) \right] \frac{d\bar{q}_6}{\bar{q}_6}
 \nonumber \\
 & = &
 \omega^{\mathrm{Kronecker},\tau}_{3,1} - 4 \omega^{\mathrm{Kronecker},\tau}_{3,2}.
\eq
Suppose we are interested in the iterated integral
\bq
\label{example_2}
 I\left(\omega_0,\omega^{\mathrm{Kronecker},\tau}_3;\bar{q}_6\right).
\eq
This integral is evaluated as follows:
\begin{verbatim}
#include <ginac/ginac.h>

int main()
{
  using namespace std;
  using namespace GiNaC;

  Digits = 30;

  ex tau_6 = I;
  ex qbar_6 = exp(2*Pi*I*tau_6);
    
  ex omega_0 = basic_log_kernel();

  ex C_3  = I/sqrt(numeric(3));
  ex g3_1 = Kronecker_dtau_kernel(3,numeric(1,3),1,C_3);
  ex g3_2 = Kronecker_dtau_kernel(3,numeric(1,3),2,C_3);
    
  ex expr = iterated_integral(lst{omega_0,g3_1},qbar_6)
    -4*iterated_integral(lst{omega_0,g3_2},qbar_6);

  cout << "I = " << expr.evalf() << endl;

  return 0;
}
\end{verbatim}
Running this program will print out
\begin{verbatim}
I = 0.001863090057835543048808657035227425650156
    -3.1647606893242120514435656176022897222E-42*I
\end{verbatim}
The imaginary part is compatible with zero within the numerical precision.
It is no coincidence that the real part agrees with
the evaluation of eq.~(\ref{example_1}) within the numerical precision.
The iterated integrals in eq.~(\ref{example_1}) and eq.~(\ref{example_2}) give the same value.

\subsubsection{Example 3}

As a third example we consider the evaluation of an elliptic multiple polylogarithm.
We consider
\bq
 \widetilde{\Gamma}\!\left({\begin{smallmatrix} 0 & 1 \\ 0 & z_2 \\ \end{smallmatrix}}; z; \tau \right)
\eq
with $z_2=1/3$, $z=1/10$ and $\tau=i$.
The elliptic multiple polylogarithm
is evaluated as follows:
\begin{verbatim}
#include <ginac/ginac.h>

int main()
{
  using namespace std;
  using namespace GiNaC;

  Digits = 30;

  ex z = numeric(1,10);

  ex tau = I;
  ex qbar = exp(2*Pi*I*tau);

  ex C_0 = pow(2*Pi*I,-2);
  ex C_1 = pow(2*Pi*I,-1);

  ex omega_1   = Kronecker_dz_kernel(1,0,tau,1,C_0);
  ex omega_2   = Kronecker_dz_kernel(2,numeric(1,3),tau,1,C_1);
    
  ex expr = iterated_integral(lst{omega_1,omega_2},z);

  cout << "I = " << expr.evalf() << endl;

  return 0;
}
\end{verbatim}
Running this program will print out
\begin{verbatim}
I = 2.94464934831952392205025799050160320365E-4
    +2.66276198352254155922945883386642082752E-43*I
\end{verbatim}
The imaginary part is compatible with zero within the numerical precision.

\subsection{$\bar{q}$-expansions}

Apart from evaluating iterated integrals it is sometimes useful to access the $\bar{q}$-expansions of modular forms.
As the space of modular forms ${\mathcal M}_k(\Gamma)$ is finite, the $\bar{q}$-expansions can be used to
express any $\eta \in {\mathcal M}_k(\Gamma)$ as a linear combination of a basis of ${\mathcal M}_k(\Gamma)$.

\subsubsection{Defining a basis for the Eisenstein subspace}

This example shows, how to define a basis for the Eisenstein subspace ${\mathcal E}_k(\Gamma_1(12))$
of ${\mathcal M}_k(\Gamma_1(12))$ for the modular weights $k=1$ and $k=2$.
Higher weights follow a similar pattern.
At level $N=12$ we have to consider only primitive Dirichlet characters which are obtained from the Kronecker symbol.
Dirichlet characters which cannot be obtained from the Kronecker symbol have a conductor, which is not a divisor of $12$.
In order to construct a basis of $\mathcal{E}_k(\Gamma_1(12))$
we have to consider the set of all possible $E_{k,12,a,b,K}(\tau)$'s such that $12$ is an integer multiple of $K |a| |b|$
and the conditions in eq.~(\ref{condition_k_eq_1}) and eq.~(\ref{condition_k_eq_1}) are satisfied.

For $\Gamma_1(12)$ the primitive Dirichlet characters $\chi_a(n)$ with $a\in\{1,-3,-4,12\}$ are relevant.
We have
\bq
 \chi_1(-1) \; = \; 1,
 \;\;\;\;\;\;
 \chi_{-3}(-1) \; = \; -1,
 \;\;\;\;\;\;
 \chi_{-4}(-1) \; = \; -1,
 \;\;\;\;\;\;
 \chi_{12}(-1) \; = \; 1.
\eq
At modular weight $1$ we must have $\chi_b(-1)=-1$ (see eq.~(\ref{condition_k_eq_1}))
and therefore $b \in \{-3,-4\}$.
As $|a||b|$ must divide $12$, the only choice for $a$ is then $a=1$.
Thus a basis of ${\mathcal E}_1(\Gamma_1(12))$ is given by
\bq
 E_{1,12,1,-3,1}\left(\tau\right),
 \;\;
 E_{1,12,1,-3,2}\left(\tau\right),
 \;\;
 E_{1,12,1,-3,4}\left(\tau\right),
 \;\;
 E_{1,12,1,-4,1}\left(\tau\right),
 \;\;
 E_{1,12,1,-4,3}\left(\tau\right).
\eq
The following code fragment computes the $\bar{q}_{12}$-expansion to order $100$:
\begin{verbatim}
symbol q("q");
cout 
  << Eisenstein_kernel(1, 12, 1, -3, 1).q_expansion_modular_form(q, 100) 
  << endl;
cout 
  << Eisenstein_kernel(1, 12, 1, -3, 2).q_expansion_modular_form(q, 100) 
  << endl;
cout 
  << Eisenstein_kernel(1, 12, 1, -3, 4).q_expansion_modular_form(q, 100) 
  << endl;
cout 
  << Eisenstein_kernel(1, 12, 1, -4, 1).q_expansion_modular_form(q, 100) 
  << endl;
cout 
  << Eisenstein_kernel(1, 12, 1, -4, 3).q_expansion_modular_form(q, 100) 
  << endl;
\end{verbatim}
This gives
\bq
 E_{1,12,1,-3,1}\left(\tau\right)
 & = &
 \frac{1}{6} + \bar{q}_{12} + \bar{q}_{12}^3 + \bar{q}_{12}^4 
 + 2 \bar{q}_{12}^7 + \bar{q}_{12}^9 + \bar{q}_{12}^{12} + \dots,
 \nonumber \\
 E_{1,12,1,-3,2}\left(\tau\right)
 & = &
 \frac{1}{6} + \bar{q}_{12}^2 + \bar{q}_{12}^6 + \bar{q}_{12}^8 
 + \dots,
 \nonumber \\
 E_{1,12,1,-3,4}\left(\tau\right)
 & = &
 \frac{1}{6} + \bar{q}_{12}^4 + \bar{q}_{12}^{12}
 + \dots,
 \nonumber \\
 E_{1,12,1,-4,1}\left(\tau\right)
 & = &
 \frac{1}{4} + \bar{q}_{12} + \bar{q}_{12}^2 + \bar{q}_{12}^4 
 + 2 \bar{q}_{12}^5 + \bar{q}_{12}^8 + \bar{q}_{12}^9 + 2 \bar{q}_{12}^{10} + \dots,
 \nonumber \\
 E_{1,12,1,-4,3}\left(\tau\right)
 & = &
 \frac{1}{4} + \bar{q}_{12}^3 + \bar{q}_{12}^6 + \bar{q}_{12}^{12} 
 + \dots.
\eq
At modular weight $k=2$ we must have 
\bq
 \chi_a\left(-1\right) \chi_b\left(-1\right) & = & 1.
\eq
This gives
\bq
 & &
 E_{2,12,1,1,2}\left(\tau\right),
 \;\;
 E_{2,12,1,1,3}\left(\tau\right),
 \;\;
 E_{2,12,1,1,4}\left(\tau\right),
 \;\;
 E_{2,12,1,1,6}\left(\tau\right),
 \;\;
 E_{2,12,1,1,12}\left(\tau\right),
 \nonumber \\
 & &
 E_{2,12,1,12,1}\left(\tau\right),
 \;\;
 E_{2,12,12,1,1}\left(\tau\right),
 \;\;
 E_{2,12,-3,-4,1}\left(\tau\right),
 \;\;
 E_{2,12,-4,-3,1}\left(\tau\right),
\eq
as a basis of ${\mathcal E}_2(\Gamma_1(12))$.
The $\bar{q}_{12}$-expansions are computed in the same way as above.

The $\bar{q}$-expansion can also be done in \verb|ginsh|. For the first example from above one would simply type
\begin{verbatim}
> q_expansion_modular_form(Eisenstein_kernel(1, 12, 1, -3, 1), q, 13);
1/6+1*q+1*q^3+1*q^4+2*q^7+1*q^9+1*q^12+Order(q^13)
\end{verbatim}

\subsubsection{Defining a cusp form}

This example shows how to define a modular form through a polynomial in Eisenstein series.
Let
\bq
 \eta_4\left(\tau\right) \; = \; E_{4,1,1,1,1}\left(\tau\right),
 & &
 \eta_6\left(\tau\right) \; = \; E_{6,1,1,1,1}\left(\tau\right)
\eq
be the two Eisenstein series of modular weight $4$ and $6$, respectively, 
which generate the modular forms of $\mathrm{SL}_2({\mathbb Z})$ as a ring.
Suppose, we would like to construct the cusp form
\bq
 \eta_{12}\left(\tau\right) & = & \left(2\pi i\right)^{12} \Delta\left(\tau\right),
\eq
with $\Delta(\tau)$ given in eq.~(\ref{def_Delta}).
From eq.~(\ref{relation_e_k_to_E_k_1_1}) we have
\bq
 \eta_{12}\left(\tau\right) & = & 8000 \eta_4\left(\tau\right)^3 - 147 \eta_6\left(\tau\right)^2.
\eq
The following code fragment defines $\eta_{12}(\tau)$ and computes the $\bar{q}$ expansion to order $100$:
\begin{verbatim}
symbol q("q");

ex eta_4 = Eisenstein_kernel(4, 1, 1, 1, 1);
ex eta_6 = Eisenstein_kernel(6, 1, 1, 1, 1);

ex P = 8000*pow(eta_4,3)-147*pow(eta_6,2);

ex eta_12 = modular_form_kernel(12, P);

cout 
  << ex_to<modular_form_kernel>(eta_12).q_expansion_modular_form(q, 100) 
  << endl;
\end{verbatim}
This gives
\bq
 \bar{q}
 -24 \bar{q}^2
 +252 \bar{q}^3
 -1472 \bar{q}^4
 +4830 \bar{q}^5
 -6048 \bar{q}^6
 + \dots
\eq

\subsection{Numerical evaluation of a Feynman integral}
\label{sect:example_Feynman_integral}

Let us consider the equal-mass sunrise integral in two space-time dimensions:
\bq
 S_{111}\left(x\right)
 & = &
 \frac{m^2}{\pi^2}
 \int d^2k_1 \int d^2k_2 \int d^2k_3
 \frac{\delta^2\left(p-k_1-k_2-k_3\right)}{\left(k_1^2-m^2\right)\left(k_1^2-m^2\right)\left(k_1^2-m^2\right)}.
\eq
This integral depends on the variable $p$ (the external momentum four-vector) 
and the variable $m$ (the internal mass)
only through the ratio $x=p^2/m^2$.
Feynman's $i\delta$-prescription translates into an infinitesimal small positive imaginary part of $x$.
The following code computes this Feynman integral
for $x \in ({\mathbb R}\cup\{\infty\})\backslash\{0,1,9,\infty\}$.
\begin{verbatim}
#include <ginac/ginac.h>

int main()
{
  using namespace std;
  using namespace GiNaC;

  Digits = 30;

  // input x = p^2/m^2, x real and not equal to {0,1,9}
  numeric x = numeric(901,100);

  numeric sqrt_3 = sqrt(numeric(3));
  numeric sqrt_x = sqrt(x);
  numeric k2     = 16*sqrt_x/pow(1+sqrt_x,numeric(3))/(3-sqrt_x);
 
  ex  pre = 4*pow(1+sqrt_x,numeric(-3,2))*pow(3-sqrt_x,numeric(-1,2));
  if (x > 9) pre = -pre;
  ex psi1 = pre*EllipticK(sqrt(k2));
  ex psi2 = pre*I*EllipticK(sqrt(1-k2));
  if ((x < 3-2*sqrt_3) || (x > 1)) psi1 += 2*psi2;
  if ((x > 1) && (x < 9)) psi1 += 2*psi2;
  ex tau  = psi2/psi1;
  ex qbar = exp(2*Pi*I*tau);

  ex L_0   = basic_log_kernel();
  ex eta_1 = Eisenstein_kernel(3, 6, -3, 1, 1);
  ex eta_2 = Eisenstein_kernel(3, 6, -3, 1, 2);
  ex E_3   = modular_form_kernel(3, eta_1-8*eta_2);

  ex Cl2  = numeric(1,2)/I*(Li(2,exp(2*Pi*I/3))-Li(2,exp(-2*Pi*I/3)));
  ex S111 = 3*psi1/Pi*(Cl2-sqrt_3*iterated_integral(lst{L_0,E_3},qbar));

  cout << "S111 = " << S111.evalf() << endl;

  return 0;
}
\end{verbatim}
The formulae underlying this code have been taken from \cite{Adams:2017ejb,Bogner:2017vim,Honemann:2018mrb}.
There is a convention how mathematical software should evaluate a function on a branch cut:
{\em implementations shall map a cut so the function is continuous as the cut is approached coming
around the finite endpoint of the cut in a counter clockwise direction} \cite{C99standard}.  
GiNaC follows this convention. 
In physics, Feynman's $i\delta$-prescription dictates how a function should be evaluated on a branch cut.
The lines
\begin{verbatim}
  if (x > 9) pre = -pre;
  if ((x > 1) && (x < 9)) psi1 += 2*psi2;
\end{verbatim}
correct for a mismatch between the standard convention for mathematical software and
Feynman's $i\delta$-prescription.
In a neighbourhood of the point $x=0$ the code is highly efficient.
In a neighbourhood of the points $x \in \{1,9,\infty\}$ the convergence is slow.
Running the code for
\bq 
 x & = & 9.01
\eq
yields
\begin{verbatim}
S111 = 13.1694380519281544350998973329994190351
       +5.69347485398690488436191383523244525036*I
\end{verbatim}
Please note that the point $x=9.01$ is a point where the convergence is slow.
Table~\ref{cputime} compares the required CPU time on a single core of a standard laptop 
for the point $x=9.01$ with the point $x=0.01$ 
(where the convergence is fast) as a function of the requested digits.
\begin{table}
\begin{center}
\begin{tabular}{|c|rrr|}
\hline
 Digits & $30$ & $100$ & $300$ \\
\hline
 $x=0.01$ & $<1\mathrm{s}$ & $<1\mathrm{s}$ & $<1\mathrm{s}$ \\
 $x=9.01$ & $1\mathrm{s}$ & $15\mathrm{s}$ & $283\mathrm{s}$ \\
\hline
\end{tabular}
\caption{\label{cputime}
CPU time in seconds to compute the equal-mass sunrise integral $S_{111}$ for a given value $x$ with a precision of $n$ digits.
}
\end{center}
\end{table}
The two points $x=9.01$ and $x=0.01$ represent two extreme cases, where the convergence is slow, respectively fast.
Please note that in the neighbourhoods of the points $x\in\{1,9,\infty\}$ alternative representations
can be used (and should be used), which provide significant faster convergence.
These alternative representations are again given as iterated integrals of modular forms \cite{Honemann:2018mrb,Duhr:2019rrs}.

\section{Advanced usage, limitations and outlook}
\label{sect:advanced}

\subsection{Truncation}

The iterated integrals are evaluated by a series expansion. 
Let us write schematically the series truncated after the order $N$ term as
\bq
 I^{\mathrm{approx}}(N)
 & = &
 \sum\limits_{i_1=1}^N d_{i_1}.
\eq
The default truncation criterion is given in eq.~(\ref{truncation_criterion}):
\bq
 I^{\mathrm{approx}}\left(N\right)
 \; \sim \;
 I^{\mathrm{approx}}\left(N-1\right)
 & \mbox{and} &
 d_N \; \neq \; 0.
\eq
The symbol $\sim$ means that the two numbers agree as floating-point numbers 
within a given numerical precision.

The reason for requiring $d_n \neq 0$ in the truncation criterion is as follows:
It occurs quite often, that some terms $d_n$ are exactly zero.
For example, this occurs for iterated integrals of depth one for the integration kernels
$\omega^{\mathrm{Kronecker},\tau}_{k,K}\left(z_j\right)$ and $\omega^{\mathrm{Eisenstein}}_{k,N,a,b,K}$, whenever $K>1$.
Without the condition $d_n \neq 0$ the series would be truncated as soon as the first term $d_n$ is zero.
For example, if the iterated integral has the $\bar{q}$-expansion
\bq
 I
 & = &
 c_0 + c_2 \bar{q}^2 + c_4 \bar{q}^4 + c_6 \bar{q}^6 + \dots,
\eq
omitting the condition $d_n \neq 0$ would return $c_0$ as numerical approximation, since $d_1=0 \cdot \bar{q}$ is zero.

There are some cases, which cannot be handled properly by the standard truncation criterion.
\begin{enumerate}
\item The first case is rather trivial. Consider the integral
\bq
 I & = &
 \int\limits_0^{x_0} x dx
\eq
with the user-defined kernel 
\bq
\label{example_user_defined}
 \omega^{\mathrm{user}}\left(f\right)
 \; = \; 
 f\left(x\right) dx,
 & &
 f\left(x\right)
 \; = \; 
 x.
\eq
The series expansion of the integrand terminates with the $x^1$-term.
Afterwards, there are no non-zero terms, and an algorithm based on eq.~(\ref{truncation_criterion}) will go into an infinite loop,
looking for a non-zero term.
\item The second case is more subtle.
As an example consider the integration kernel
$\omega^{\mathrm{Eisenstein,h}}_{k,N,r,s}$. The $\bar{q}$-expansion of the integrand involves roots of unity.
Consider the combination
\bq
\label{numerical_counter_example}
\lefteqn{
 \exp\left(2\pi i \frac{1}{6}\right)
 +
 \exp\left(2\pi i \frac{2}{6}\right)
 +
 \exp\left(2\pi i \frac{4}{6}\right)
 +
 \exp\left(2\pi i \frac{5}{6}\right)
 = } & &
 \\
 & = &
  \left(\frac{1}{2}+\frac{i}{2}\sqrt{3}\right)
 +
  \left(-\frac{1}{2}+\frac{i}{2}\sqrt{3}\right)
 +
  \left(-\frac{1}{2}-\frac{i}{2}\sqrt{3}\right)
 +
  \left(\frac{1}{2}-\frac{i}{2}\sqrt{3}\right)
 \; = \; 0.
 \nonumber
\eq
This is zero. However, the roots are computed numerically as floating-point numbers and it may happen that the result
is a small non-zero number at the order of the numerical precision.
Usually a term like in eq.~(\ref{numerical_counter_example}) is multiplied by prefactors and it may happen that $d_N$ is non-zero but small, such that
adding $d_N$ to $I^{\mathrm{approx}}(N-1)$ will not change the floating-point representation of $I^{\mathrm{approx}}(N-1)$.
As $d_N \neq 0$ and $I^{\mathrm{approx}}(N-1) \sim I^{\mathrm{approx}}(N)$ the default truncation criterion will truncate the series incorrectly.
\end{enumerate}
Both cases can be handled by giving a third argument \verb|N_trunc| to \verb|iterated_integral|, which specifies that the series should be summed up to
$N_{\mathrm{trunc}}$.
The following short program integrates the function $f(x)=x$ from zero to one:
\begin{verbatim}
#include <ginac/ginac.h>

int main()
{
  using namespace std;
  using namespace GiNaC;

  Digits = 30;

  symbol x("x");
  ex f = x;
  ex omega = user_defined_kernel(f,x);
  ex expr = iterated_integral(lst{omega},numeric(1),10);

  cout << "I = " << expr.evalf() << endl;

  return 0;
}
\end{verbatim}
Some advice to the user: When relying on the default truncation criteria, it is always a good idea
to cross check the results with a suitably specified truncation parameter.
In particular it is recommended to use an explicit truncation parameter
\begin{itemize}
\item if for all integration kernels the series expansion in the integration variable terminates
after a finite number of terms
(this may happen for user-supplied integration kernels by using the class \verb|user_defined_kernel(f, y)|),
\item if integration kernels from the class \verb|Eisenstein_h_kernel(k, N, r, s, C_k)| with $N \notin \{1,2,4\}$ are used.
\end{itemize}
Usually the need for an explicit truncation parameter is already detected early on, 
i.e. typically already for iterated integrals of depth one.

\subsection{Limitations and outlook}

The current implementation is limited to the case, where
all integrands have a convergent Laurent series expansion with at most
a simple pole at the base point of the integration.
It does not include algorithms to speed up the computation for points close to the boundary of the region of convergence
nor does it include algorithms to continue the integration path beyond the region of convergence.
For specific subclasses of the iterated integrals discussed in this paper 
there are methods which address these points \cite{Passarino:2017EPJC,Duhr:2019rrs}.
While it is certainly desirable to have such methods, we would also like to emphasise
that the tools of the current implementation offer a satisfactory support, if the calculation is organised
in an appropriate way:

Let us first discuss the issue of analytic continuation, i.e. the continuation
of the integration path beyond the region of convergence of the series expansion of the integrand.
This problem is known from multiple polylogarithms and occurs when $|y|$ exceeds $|z_j|$ in eq.~(\ref{def_omega_mpl}).
The same issue arises in the $\tilde{\Gamma}$-functions when $|z|$ exceeds $|z_j|$ in eq.~(\ref{def_omega_Kronecker_z})
for $k=2$.
The problem is absent for iterated integrals, which are integrations in $\tau$ (or $\bar{q}$):
Here, we integrate from $\bar{q}=0$ to $|\bar{q}|<1$ and the integrands are holomorphic
on the punctured unit disc $0<|\bar{q}|<1$.
At the puncture $\bar{q}=0$ the integrands may have at most a simple pole.
If we compute the Feynman integrals from their differential equations and if we change the kinematic variables
to the standard variables $(z_1,\dots,z_{n-1},\tau)$ for the moduli space ${\mathcal M}_{1,n}$
we may integrate the differential equation either in a $z$-variable (yielding $\tilde{\Gamma}$-functions)
or in $\tau$ (yielding iterated integrals with integrations in $\tau$, or equivalently $\bar{q}$).
The issue of analytic continuation can be avoided, if one chooses to integrate the differential equation in $\tau$.
The point $\tau=i\infty$ (corresponding to $\bar{q}=0$) is a convenient boundary point.
The elliptic curve degenerates at this point and the Feynman integrals can usually be expressed at this point
in terms of multiple polylogarithms in the remaining variables.
We have already seen an example in section~\ref{sect:example_Feynman_integral}:
The $\bar{q}$-expansion of the sunrise integral in this example is an expansion around $\bar{q}=0$, corresponding
to an expansion around $p^2=0$. Nevertheless, the $\bar{q}$-expansion computes correctly the value of the
sunrise integral above the threshold $p^2>9m^2$.

Let us now turn to the second issue: How to avoid a slow convergence of the iterated integral, if the iterated
integral is integrated up to $|\bar{q}| \lesssim 1$.
The method of choice is to perform a modular transformation
\bq
 \tau' & = & \frac{a\tau+b}{c\tau+d},
 \;\;\;\;\;\;
 \left( \begin{array}{cc} a & b \\ c & d \\ \end{array} \right)
 \; \in \;
 \mathrm{SL}_2({\mathbb Z})
\eq
such that $|\bar{q}'| < |\bar{q}|$
and to re-express the transformed differential one-forms in the differential equation for the Feynman integral
again as a linear combination from the classes
\bq
 \omega^{\mathrm{Kronecker},\tau}_{k,K}\left(z_j\right),
 \;\;\;
 \omega^{\mathrm{Eisenstein}}_{k,N,a,b,K},
 \;\;\;
 \omega^{\mathrm{Eisenstein,h}}_{k,N,r,s},
 \;\;\;
 \omega^{\mathrm{modular}}\left(P_k\right).
\eq
It is usually more efficient to do this at the level of the differential equation for the Feynman integral and not
at the level of the final linear combination of iterated integrals.
The reason is that only a limited number of different 
differential one-forms appear in the differential equation.
Each differential one-form needs to be transformed only once, and this can be done efficiently at the level
of the differential equation. 
At the level of linear combinations of iterated integrals one would transform the same 
differential one-form over and over again.
A worked out example where this technique has been applied can be found in refs.~\cite{Honemann:2018mrb,Duhr:2019rrs}.

\section{Conclusions}
\label{sect:conclusions}

In this paper we reported on numerical evaluation methods for iterated integrals related 
to elliptic Feynman integrals.
The methods allow to evaluate these iterated integrals
to arbitrary precision within the region of convergence of the series expansion of the integrand.
All routines are integrated in the computer algebra package GiNaC 
and can be obtained by downloading this library \cite{Bauer:2000cp}.

\subsection*{Acknowledgements}

We would like to thank Claude Duhr and Lorenzo Tancredi for useful discussions.


\begin{appendix}

\section{Notation for standard mathematical functions}
\label{appendix:standard_functions}

As notations for standard mathematical functions differ slightly in the literature, we list here the definitions and
the conventions which we follow.

The complete elliptic integral of the first kind is defined by
\bq
 K\left(k\right)
 & = &
 \int\limits_0^1 \frac{dt}{\sqrt{\left(1-t^2\right)\left(1-k^2t^2\right)}}.
\eq
The complete elliptic integral of the second kind is defined by
\bq
 E(k) & = &
 \int\limits_0^1 dt \frac{\sqrt{1-k^2 t^2}}{\sqrt{1-t^2}}.
\eq
Dedekind's eta function is defined by
\bq
 \eta\left(\tau\right) 
 & = & 
 e^{\frac{i \pi \tau}{12}} \prod\limits_{n=1}^{\infty} \left(1-e^{2\pi i n \tau}\right).
\eq
The theta functions are defined by
\bq
\theta_1\left(z,q\right) 
 & = &
 -i \sum\limits_{n=-\infty}^\infty \left(-1\right)^n q^{\left(n+\frac{1}{2}\right)^2} e^{i\left(2n+1\right)z},
 \nonumber \\
\theta_2\left(z,q\right) 
 & = &
 \sum\limits_{n=-\infty}^\infty q^{\left(n+\frac{1}{2}\right)^2} e^{i\left(2n+1\right)z},
 \nonumber \\
\theta_3\left(z,q\right) 
 & = &
 \sum\limits_{n=-\infty}^\infty q^{n^2} e^{2 i n z},
 \nonumber \\
\theta_4\left(z,q\right) 
 & = &
 \sum\limits_{n=-\infty}^\infty \left(-1\right)^n q^{n^2} e^{2 i n z}.
\eq

\section{The arithmetic-geometric mean}
\label{appendix:agm}

In this appendix we review the numerical evaluation of
the complete elliptic integral of the first and the second kind 
with the help of the arithmetic-geometric mean.
Let $a_0$ and $b_0$ be two complex numbers. For $n \in {\mathbb N}_0$ one sets
\bq
 a_{n+1} \; = \; \frac{1}{2} \left( a_n+b_n \right),
 & &
 b_{n+1} \; = \; \pm \sqrt{a_n b_n}.
\eq
The sign of the square root is chosen such that \cite{Cox:1984} 
\bq
 \left| a_{n+1} - b_{n+1} \right| & \le & \left| a_{n+1} + b_{n+1} \right|,
\eq
and in case of equality one demands in addition
\bq
 \mathrm{Im}\left( \frac{b_{n+1}}{a_{n+1}} \right) & > & 0.
\eq
The sequences $(a_n)$ and $(b_n)$ converge to a common limit 
\bq
 \lim\limits_{n \rightarrow \infty} a_n
 \; = \;
 \lim\limits_{n \rightarrow \infty} b_n
 \; = \; 
 \mathrm{agm}(a_0,b_0),
\eq
known as the arithmetic-geometric mean.
The complete elliptic integral of the first kind is given by
\bq
 K\left(k\right)
 & = &
 \frac{\pi}{2 \; \mathrm{agm}\left(1,k'\right)},
 \;\;\;
 k' 
 \; = \;
 \sqrt{1-k^2}.
\eq
For the complete elliptic integral of the second kind let us set in addition
$c_0=\sqrt{a_0^2-b_0^2}$ and
\bq
 c_{n+1}
 & = &
 \frac{c_n^2}{4 a_{n+1}}.
\eq
The complete elliptic integral of the second kind is given by
\bq
 E\left(k\right)
 & = &
 K\left(k\right)
 \left( a_0^2 - \sum\limits_{n=0}^\infty 2^{n-1} c_n^2 \right)
\eq
with initial values $a_0=1$, $b_0=k'$ and $c_0=k$.

\section{The Kronecker symbol}
\label{appendix:Kronecker_symbol}

Let $a$ be an integer and $n$ a non-zero integer with prime factorisation
$n = u p_1^{\alpha_1} p_2^{\alpha_2} ... p_k^{\alpha_k}$,
where $u \in \{1,-1\}$ is a unit.
The Kronecker symbol is defined by
\bq
 \left( \frac{a}{n} \right)
 & = & 
 \left( \frac{a}{u} \right)
 \left( \frac{a}{p_1} \right)^{\alpha_1}
 \left( \frac{a}{p_2} \right)^{\alpha_2}
 ...
 \left( \frac{a}{p_k} \right)^{\alpha_k}.
\eq
The individual factors are defined as follows:
For a unit $u$ we define
\bq
 \left( \frac{a}{u} \right)
 & = &
 \left\{ \begin{array}{rl}
 1, & u=1, \\
 1, & u=-1, \; a \ge 0, \\
 -1, & u=-1, \; a<0. \\
 \end{array} \right.
\eq
For $p=2$ we define
\bq
 \left( \frac{a}{2} \right)
 & = &
 \left\{ \begin{array}{rl}
 1, & a \equiv \pm 1 \mod 8, \\
 -1, & a \equiv \pm 3 \mod 8, \\
 0, & a \;\; \mbox{even}.  \\
 \end{array} \right.
\eq
For an odd prime $p$ we have
\bq
 \left( \frac{a}{p} \right)
 & = & 
 a^{\frac{p-1}{2}} \mod p
 \;\; = \;\;
 \left\{ \begin{array}{rl}
 1, & a \equiv b^2 \mod p, \\
 -1, & a \not\equiv b^2 \mod p, \\
 0, & a \equiv 0 \mod p. \\
 \end{array} \right.
\eq
We further set
\bq
 \left( \frac{a}{0} \right)
 & = &
 \left\{ \begin{array}{rl}
 1, & a = \pm 1 \\
 0, & \mbox{otherwise}.  \\
 \end{array} \right.
\eq
For any non-zero integer $a$ the mapping 
\bq
 n & \rightarrow & 
 \left( \frac{a}{n} \right)
\eq
is a Dirichlet character.
If $a$ is the discriminant of a quadratic field, then it is a primitive 
Dirichlet character with conductor $|a|$.
One may give a condition for $a$ being the discriminant of a quadratic field \cite{Miyake}.
We first set for $p$ being a prime number, $-1$ or $-2$
\bq
 p^\ast
 & = &
 \left\{ \begin{array}{rl}
 p, & \mbox{if} \quad p \equiv 1 \mod 4, \\
 -p, & \mbox{if} \quad p \equiv -1 \mod 4 \quad \mbox{and} \quad p \neq -1, \\
 -4, & \mbox{if} \quad p = -1, \\
 8, & \mbox{if} \quad p = 2, \\
-8, & \mbox{if} \quad p = -2. \\
 \end{array} \right.
\eq
Then an integer $a$ is the discriminant of a quadratic field if and only if
$a$ is a product of distinct $p^\ast$'s.

Including the trivial character (for which $a=1$) the possible values for $a$ with smallest absolute value
are
\bq
 1, -3, -4, 5, -7, 8, -8, -11, 12, \dots
\eq

\end{appendix}

{\footnotesize
\bibliography{/home/stefanw/notes/biblio}
\bibliographystyle{/home/stefanw/latex-style/h-physrev5}
}

\end{document}